\documentclass[arXiv,
]{actapoly}

\usepackage[utf8]{inputenc}
\usepackage{xspace}
\usepackage{aas_macros}

\newcommand\software[1]{\textsc{#1}}

\begin{document}

\title[Simple Transient Detection for the Web]
{STDWeb: Simple Transient Detection pipeline for the Web}

\correspondingauthor[S. Karpov]{Sergey Karpov}{my}{karpov@fzu.cz}

\institution{my}{Institute of Physics of the Czech Academy of Sciences,
Na Slovance 1999/2, 182 00 Prague 8, Czech Republic}

\begin{abstract}
We present a simple web-based tool, STDWeb, for a quick-look photometry and transient detection in astronomical images. It  tries to implement a self-consistent and mostly automatic data analysis workflow that would work on any image uploaded to it, allowing to perform basic interactive masking, do object detection, astrometrically calibrate the image, and build the photometric solution based on a selection of catalogues and supported filters, optionally including the color term and positionally varying zero point. It also allows you to do image subtraction using either user-provided or automatically downloaded template images, and do a  forced photometry for a specified target in either original or difference images, as well as transient detection with basic rejection of artefacts. The tool may be easily deployed allowing its integration into the infrastructure of robotic telescopes or data archives for effortless analysis of their images.
\end{abstract}


\maketitle

\section{Introduction}

The study of energetic astronomical phenomena, such as supernovae, gamma-ray bursts, or electromagnetic counterparts of gravitational wave events,
relies heavily on the ability to detect and analyze transient objects in the vast amount of data produced by modern large-scale sky surveys, as well as smaller-scale follow-up observations.
While sky survey experiments usually have dedicated photometric and transient detection pipelines\citep{iptf,ztf,ps1pipeline,goto} tailored specifically for their data and regime of operations, the observations made with ``normal'' telescopes, especially smaller ones, in follow-up regime often lack such telescope-specific data analysis. This is especially important for the projects involving data from multiple telescopes at the same time, like collaborative telescope networks\footnote{Example of such project actively using \software{STDWeb} is GRANDMA network \citep{grandmao41,grandma_ztf,grandmao42}.} or observational efforts involving amateur astronomers\footnote{Example of such professional-amateur observational partnership using \software{STDWeb} is Kilonova-Catcher project \citep{kncatcher}}. 

Uniform processing of data from heterogeneous sources (e.g. telescope networks) requires creation of analysis tools that allow both high level of automatization, configurability and overall quality control (visual and user-friendly) on all steps of the pipeline. While the building blocks, both low- and higher-level --  like routines in Image Reduction and Analysis Facility\citep{iraf} (\software{IRAF}), \software{DAOPHOT}\citep{daophot} \software{SExtractor}\citep{sextractor} or modern Python packages like \software{AstroPy}\citep{astropy} and \software{photutils}\citep{photutils} -- 
are readily available for most of required tasks, their integration into reliable pipeline is still a non-trivial effort requiring both understanding of photometry and programming affinity.

For the specific needs of image processing for GRANDMA telescope network\citep{GRANDMAO3A,GRANDMA03B,grandmao41,grandma_ztf,grandmao42} we initially created a high-level Python library, \software{STDPipe}\citep{stdpipe}, that aimed towards combining a lot of these building blocks into a set of high-level routines suitable for rapid construction of telescope-specific pipelines, but it turned out insufficient for handling really heterogeneous data in a controllable and reproducible way. So, highly motivated by the success of popular web-based \software{Astrometry.Net} service\citep{astrometrynet}, we decided to try similar approach of ``heterogeneous data processing portal'' for photometry and transient detection tasks.

As a first step towards this goal, we created \software{STDWeb} -- simple web-based tool for a quick-look photometry and transient detection in astronomical images, that may be either locally deployed, or integrated into the infrastructure of data archive or robotic telescope. Here we describe the details of its implementation and workflow (Section~\ref{sec:workflow}), along with specific details about astrometric and photometric calibration (Section~\ref{sec:calibration}) and transient detection (Section~\ref{sec:transients}). We also outline some directions of its further development in Section~\ref{sec:conclusions}.

\section{Implementation and workflow}
\label{sec:workflow}

\begin{figure*}
\centering
\includegraphics[width=0.32\linewidth]{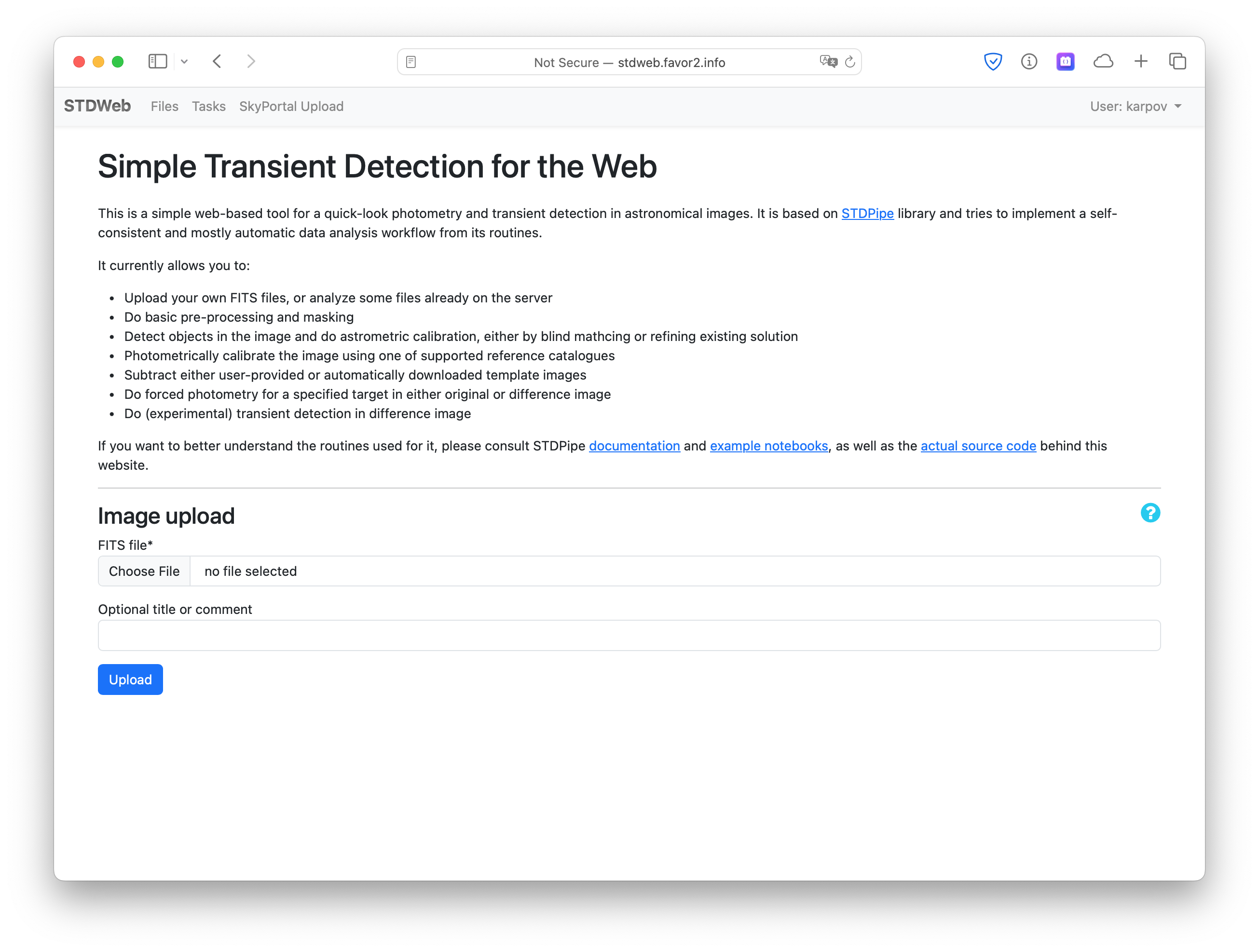}
\includegraphics[width=0.32\linewidth]{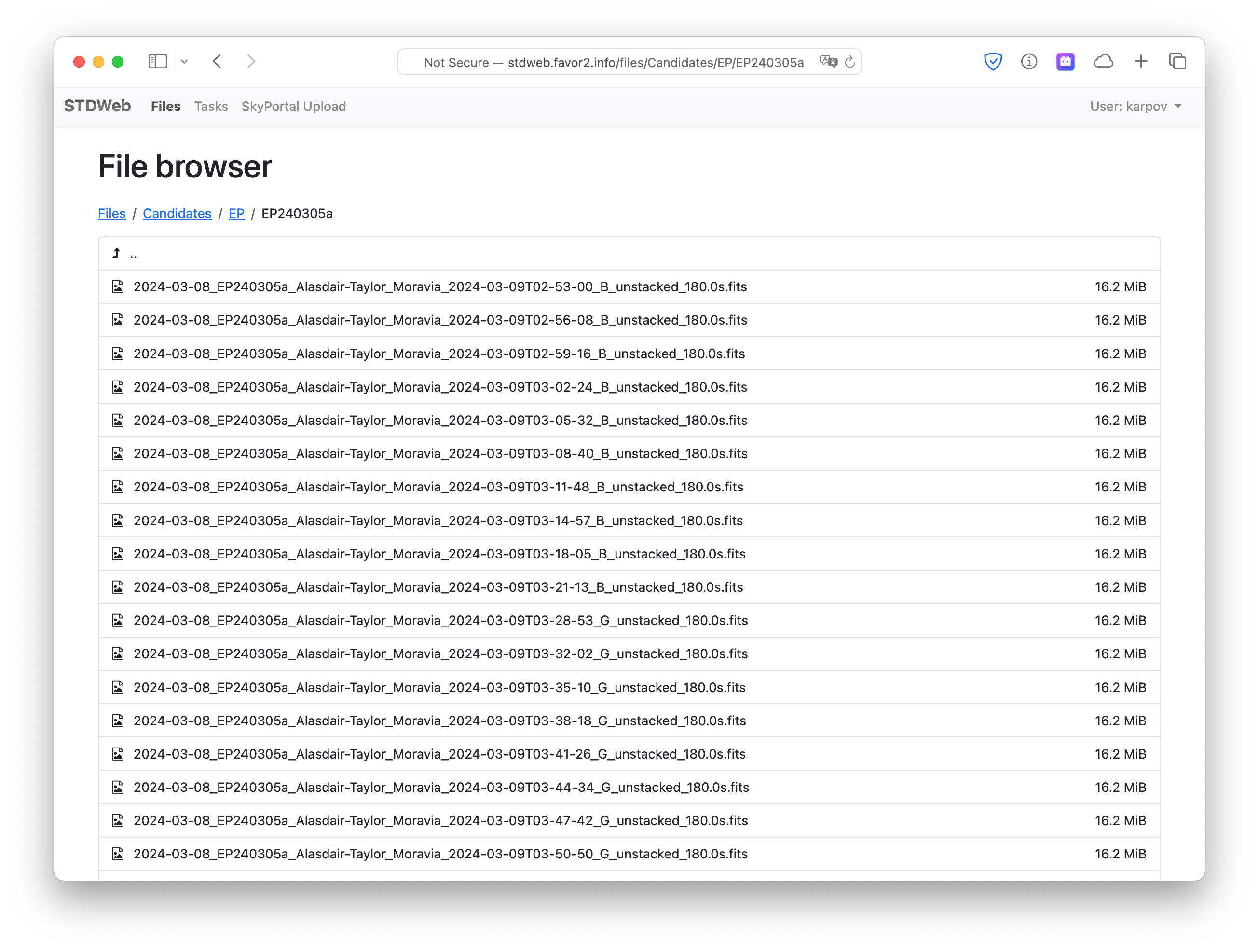}
\includegraphics[width=0.32\linewidth]{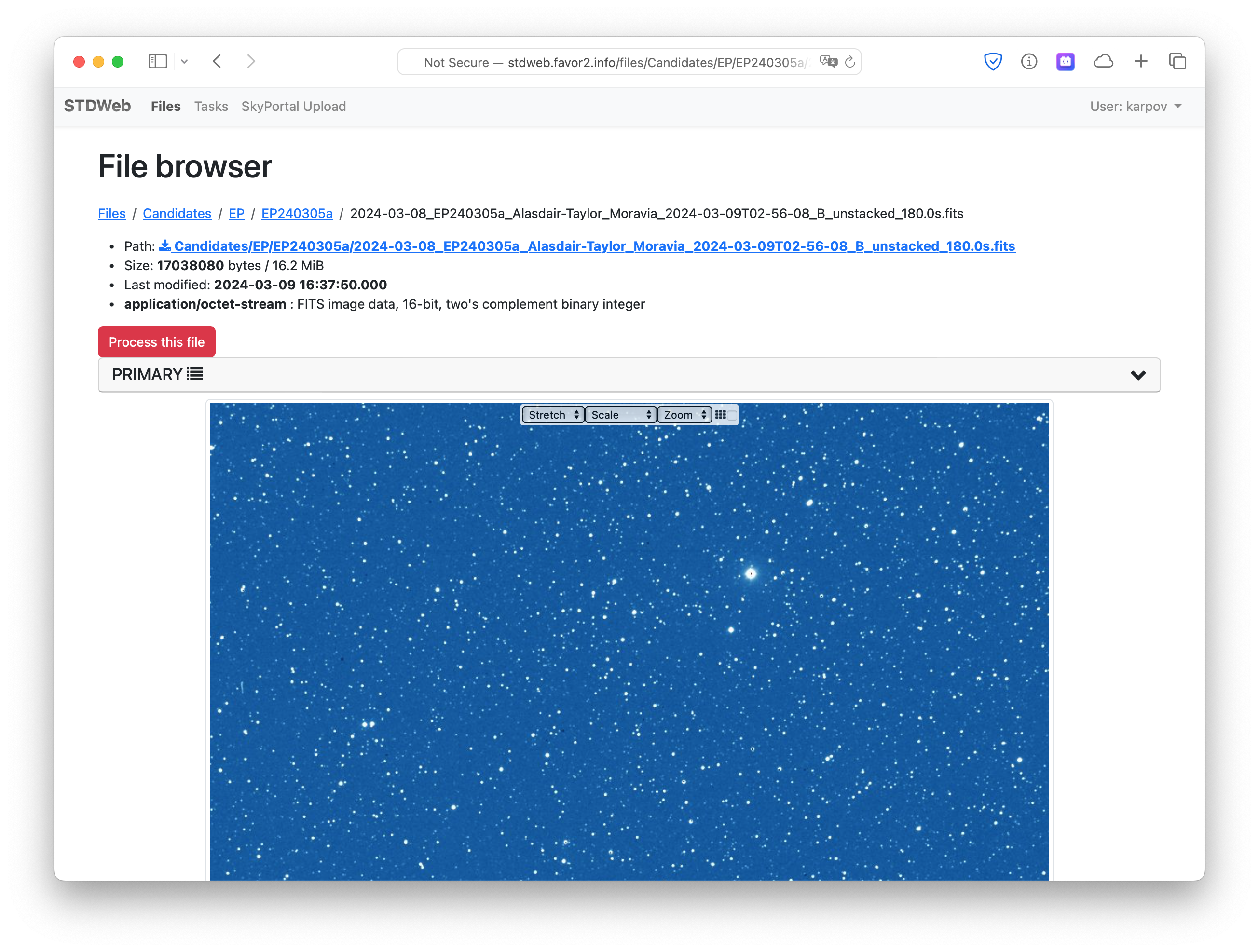}
\includegraphics[width=0.32\linewidth]{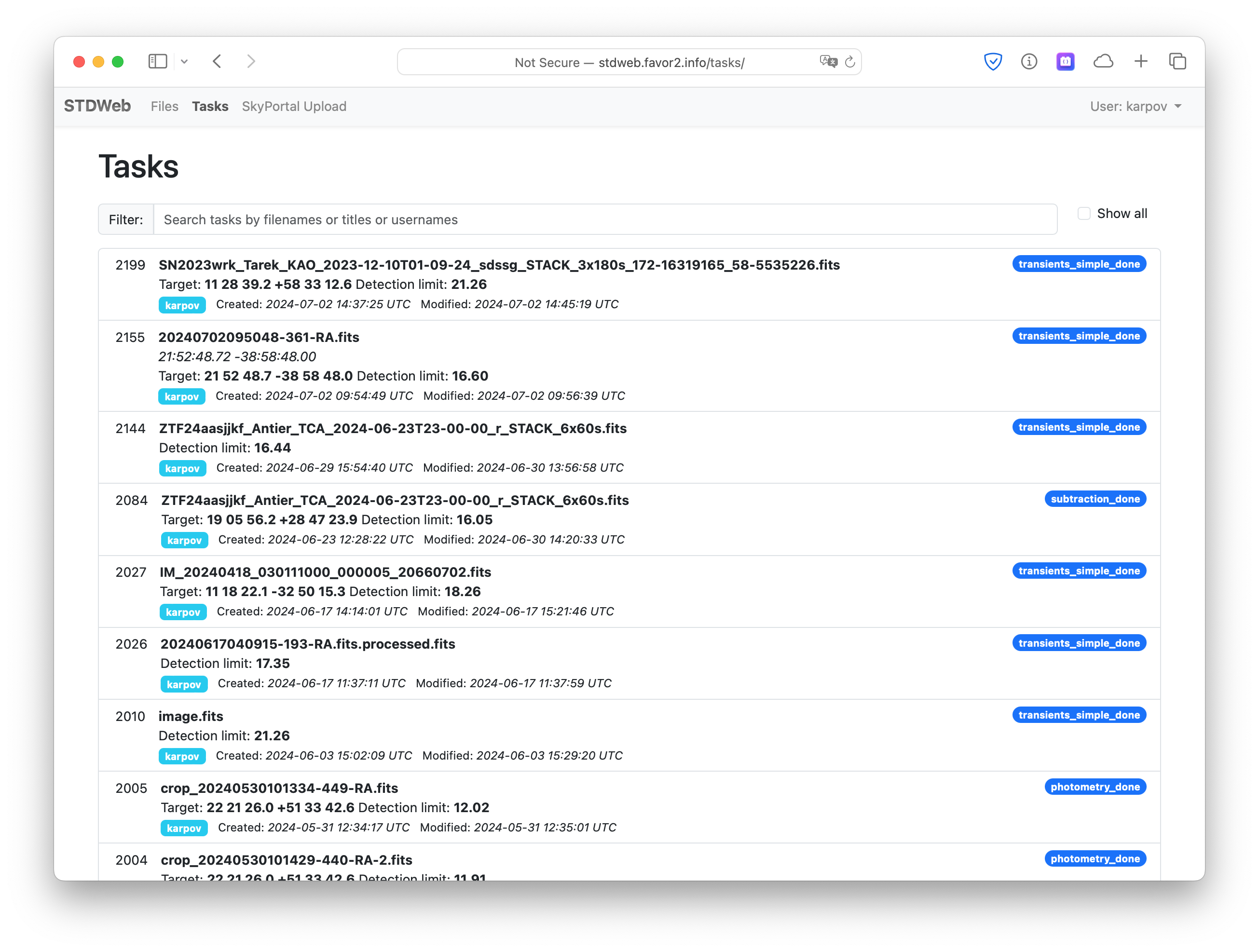}
\includegraphics[width=0.32\linewidth]{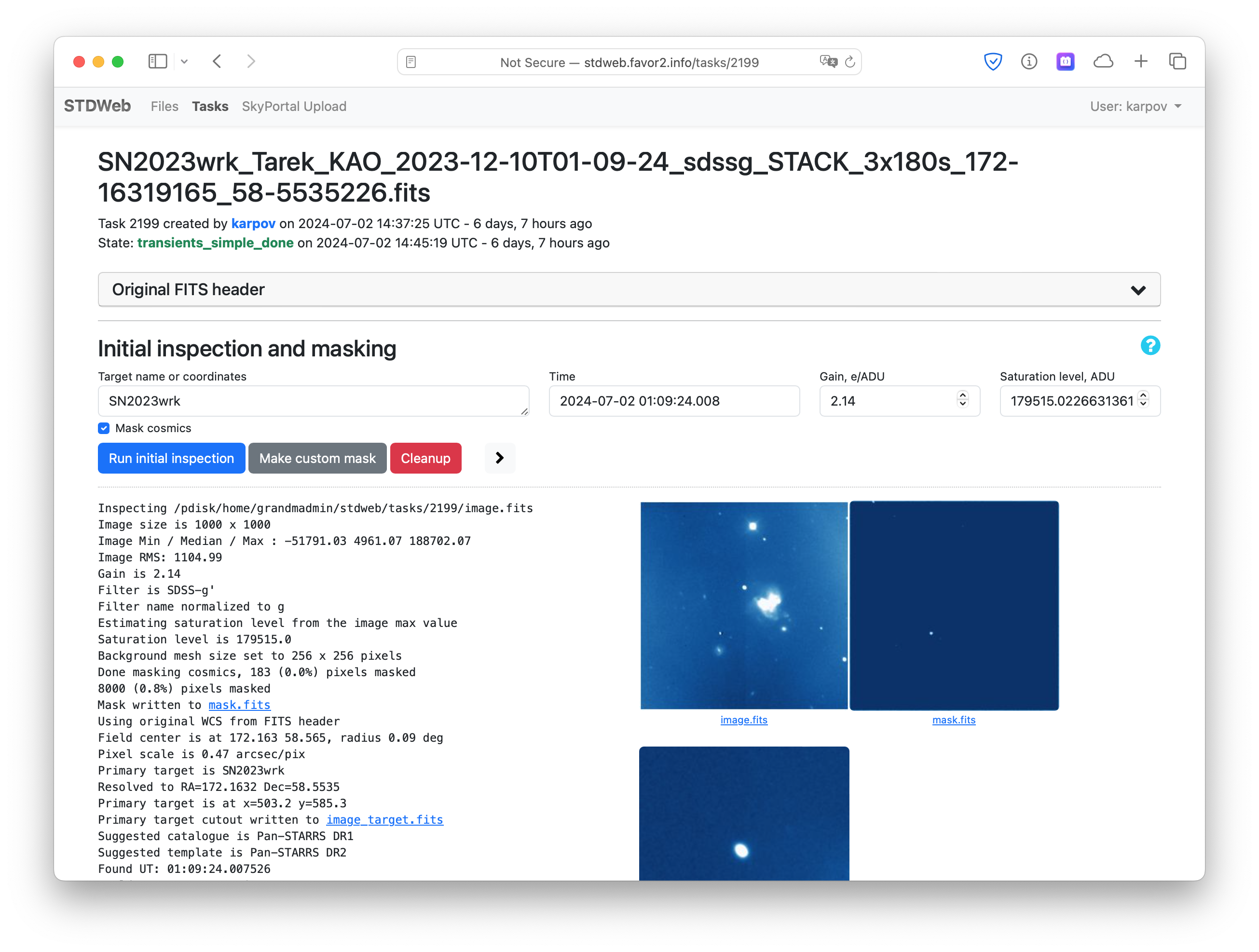}
\includegraphics[width=0.32\linewidth]{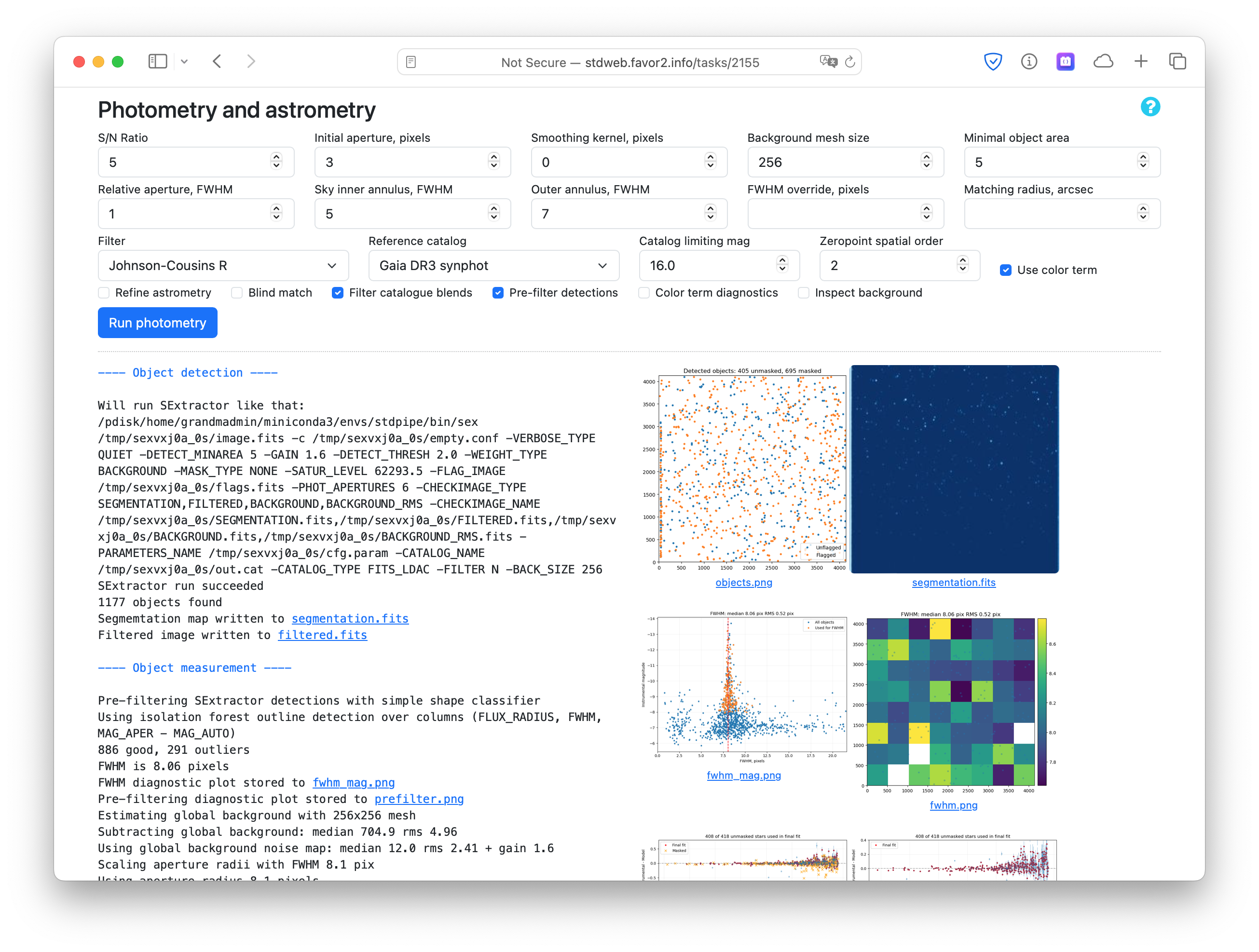}
\caption{Overall user interface of \software{STDWeb}. The panels show the image upload page (upper left panel), built-in file browser for accessing local images (upper middle), file viewer displaying local FITS image (upper right), the list of previously created tasks (lower left), task page with the options for initial inspection, textual output and diagnostic images (lower middle), and the same showing the section with options for object detection, astrometric and photometric calibration (lower right panel).} 
\label{fig:ui}
\end{figure*}

\begin{table}
\centering
\begin{tabular}{llr}
\toprule
\bfseries Name & \bfseries Purpose & \bfseries Ref.
\\ \Midrule
\multicolumn{3}{l}{External tools}
\\ \Midrule
SExtractor & Object detection & \citep{sextractor} 
\\
Astrometry.Net & Astrometric solution & \citep{astrometrynet} 
\\
SCAMP & Astrometric refinement & \citep{scamp} 
\\
SWarp & Image re-projection & \citep{swarp}
\\
HOTPANTS & Image subtraction & \citep{hotpants}
\\
\\
\multicolumn{3}{l}{Python libraries}
\\ \Midrule
Astropy & Overall infrastructure & \citep{astropy}
\\
Matplotlib & Visualization & \citep{matplotlib}
\\
Astro-SCRAPPY & Cosmic ray masking & \citep{astroscrappy}
\\
Photutils & Aperture photometry & \citep{photutils}
\\
Astroquery & Catalogue access & \citep{astroquery}
\\
Reproject & Image re-projection & \citep{reproject}
\\
Scikit-Learn & Machine Learning & \citep{scikit-learn}
\\
Django & Web interface &
\\
Celery & Backend processing &
\\\bottomrule
\end{tabular}
\caption{External software and libraries that \software{STDWeb} uses at various steps of its operation. Upper half of the list contains external binaries that should be installed separately while deploying the application, while lower part lists Python libraries that are installed automatically during the deployment.}
\label{tab:dependencies}
\end{table}

\begin{figure}
\centering
\includegraphics[width=1.0\linewidth]{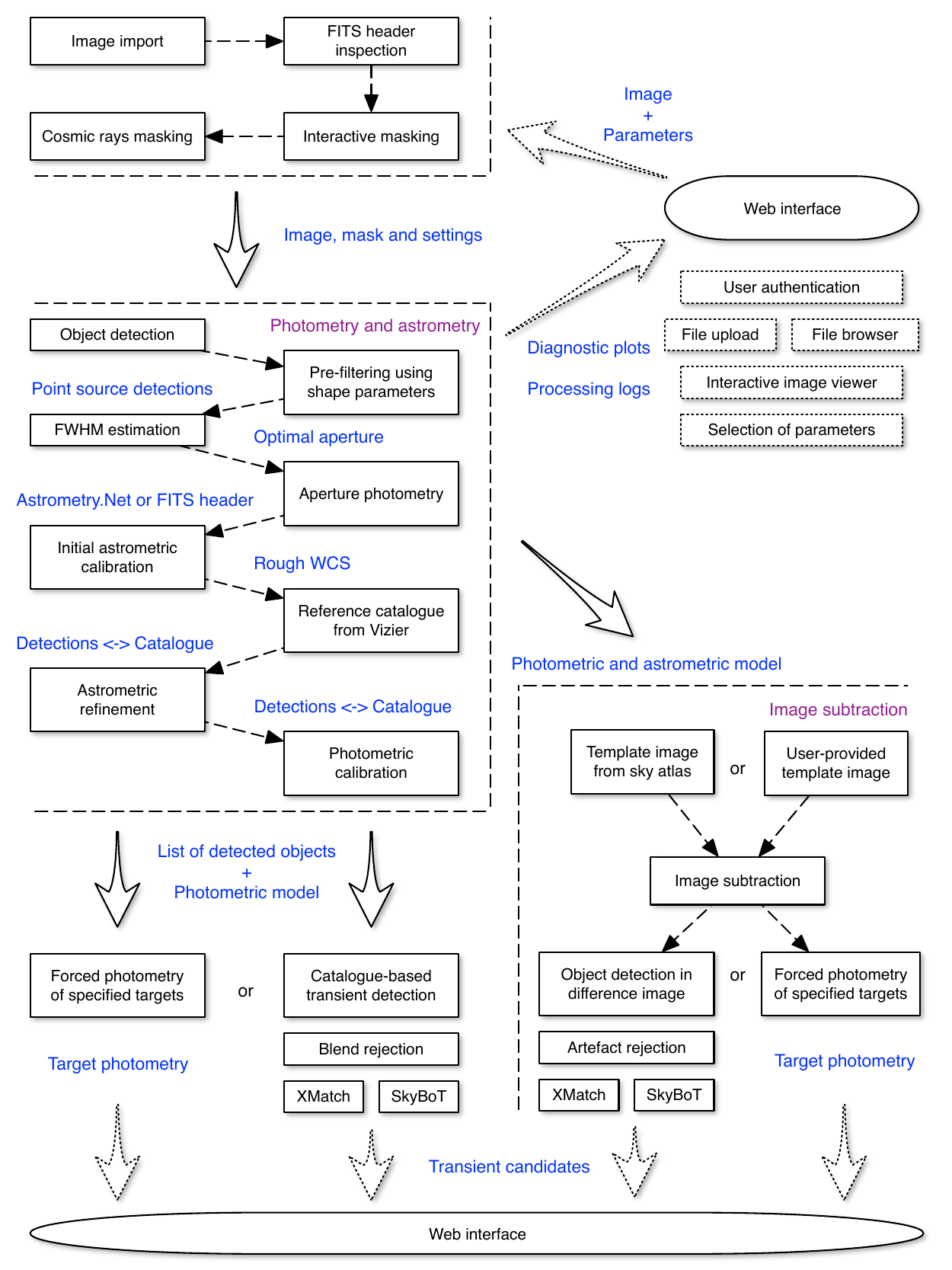}
\caption{Flowchart of the image processing steps implemented in \software{STDWeb}.}
\label{fig:flowchart}
\end{figure}

\software{STDWeb}\footnote{The code of \software{STDWeb} is available at \url{https://github.com/karpov-sv/stdweb}.}
is implemented as a \software{Django}\footnote{\url{https://www.djangoproject.com}} server-side application that handles data uploading and presentation, user authorization and interactions, as well as presentation of processing results. Actual data processing is handled in the background by \software{Celery}\footnote{\url{https://docs.celeryq.dev/}} distributed task queue that runs custom Python code based on \software{STDPipe}\citep{stdpipe} library. 

\software{STDPipe}\footnote{\software{STDPipe} is available at \url{https://github.com/karpov-sv/stdpipe} and is documented at \url{https://stdpipe.readthedocs.io/}}  is a set of Python routines for astrometry, photometry and transient detection related tasks, intended for quick and easy implementation of custom pipelines, as well as for interactive data analysis, and implemented as a library that tries to both wrap in pythonic way the access to common command-line tools like \software{SExtractor}\citep{sextractor}, \software{SCAMP}\citep{scamp}, \software{SWarp}\citep{swarp}, \software{HOTPANTS}\citep{hotpants} or \software{Astrometry.Net}\citep{astrometrynet},
and to provide extensive set of higher-level routines e.g. for aperture photometry, photometric calibration or transient candidate filtering. \software{STDWeb} uses these routines for performing its operation, while adding a set of heuristics on top of that for guessing best parameters for various steps, as well as implements some routines that are specific for its operation and not yet included in the \software{STDPipe}. An overview of external tools and libraries used by \software{STDWeb} is presented in Table~\ref{tab:dependencies}.

\software{STDWeb} allows the user either to upload the image in FITS format through web browser, or to select one from local storage (e.g. telescope or project-specific data archive available on the file system) using built-in file browser. Upon importing, a dedicated task is created for the image that will hold all configuration, intermediate processing results and operation logs related to it. The task may be re-visited at any time later in order to either review the results or initiate image re-processing.
The task page contains several sub-sections corresponding to different steps of image processing, with each section having a set of controls for adjusting the parameters relevant for that step, textual log of its execution containing various diagnostic outputs like details of running external commands or links to output files, and a set of diagnostic images and plots, each available for better viewing in a pop-up window or downloading. Pop-ups for FITS images specifically support adjusting the pixel value scaling and stretching, zooming in, and overplotting of coordinate grids, positions of user-specified targets, detected objects and catalogue stars. 
Overall view of the user interface with its various pages is shown in Figure~\ref{fig:ui}. Figure~\ref{fig:flowchart} shows an overview of the image processing steps which are discussed in more details below.


\section{Image calibration}
\label{sec:calibration}

Prior to transient detection, the image -- we assume it to be science-ready, i.e. already passed through the instrument-specific pre-processing like bias and dark current removal, flat-fielding, etc -- has to be properly masked, astrometrically and photometrically calibrated. These steps are described in detail below.

\subsection{Initial image inspection and masking}
\label{sec:inspection}

The software will automatically analyze the image trying to extract relevant information from its FITS header using a set of common keywords that define the image timestamp, its saturation level, gain settings, etc. All these settings may also be directly provided by the user, if the information is absent in the header, or is wrong -- the software performs some basic checks of whether the values are compatible with actual data range of the pixel values, in order to detect stacked or re-scaled images, and informs the user if it detects some potential problem with it. 
The software also tries to sanitize the astrometric information in the header, removing or modifying some keywords produced by popular software that break \software{AstroPy} WCS module.

Taking into account saturation level and gain, the software then creates a mask of pixels to be excluded from the consequent analysis, corresponding to saturated pixels and cosmic ray hits. For detecting the latters, \software{AstroSCRAPPY} \citep{astroscrappy} code implementing the original LACosmic \citep{lacosmic} algorithm is used. As we make no prior assumptions about the pre-processing already applied to the image (e.g. was it background-subtracted or not), the software automatically constructs the noise model to be used for these routines based on empirically estimated background noise of the image plus Poissonian noise contribution for the sources above the background.
In cases the algorithm misbehaves and starts masking normal stars as well (e.g. for undersampled images) this step may be disabled by the user. Moreover, the user may interactively create an additional ``custom mask'' to be applied on top of automatically created one, in order to mask e.g. unusable regions of the image that cannot be easily detected by the software (heavily vignetted regions, overscans, imaging artefacts or significant reflections). 

On this step the user may also provide a list of targets, either as coordinate strings in common formats or as Simbad\citep{simbad} or TNS\citep{tns} resolvable names, to be used for forced photometry later.

\subsection{Object detection}
\label{sec:detection}

\begin{figure*}
\centering
\includegraphics[width=0.49\linewidth]{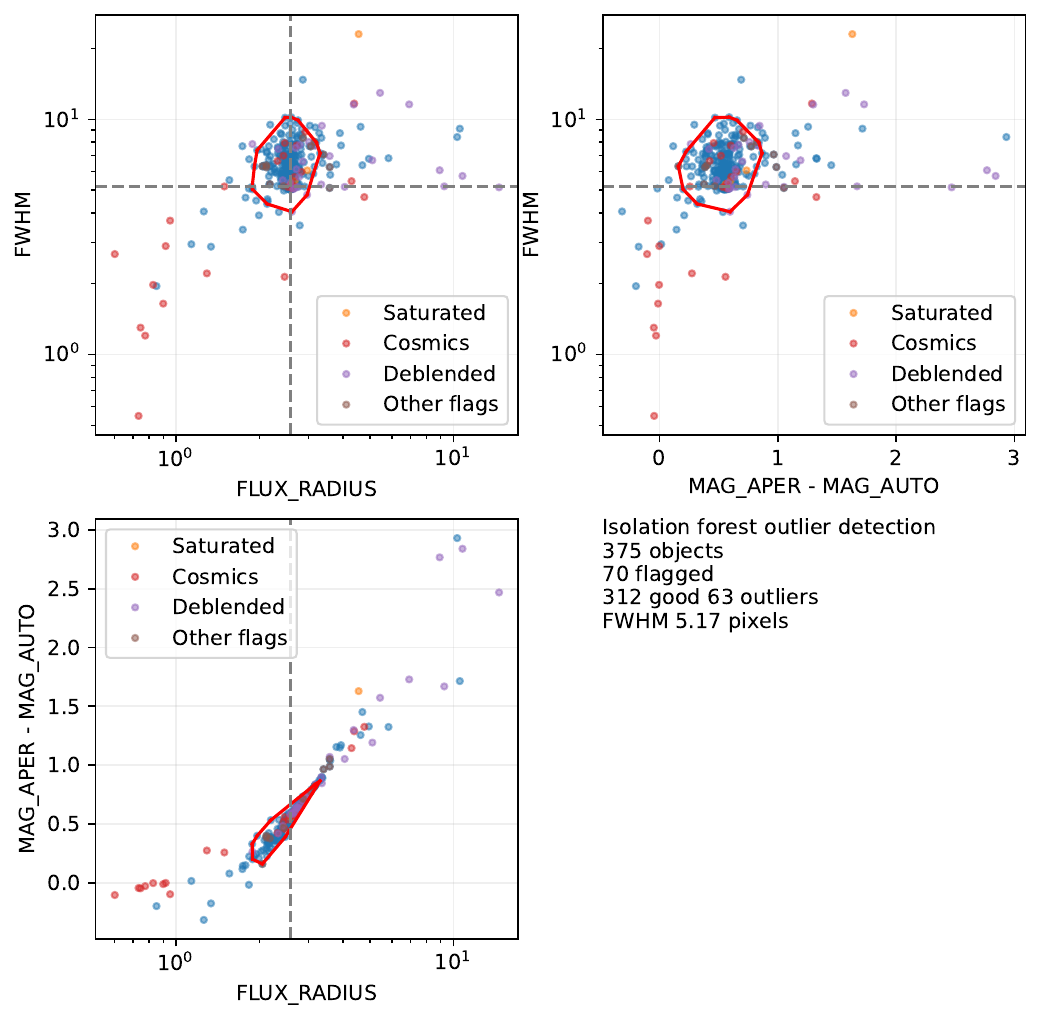}
\includegraphics[width=0.49\linewidth]{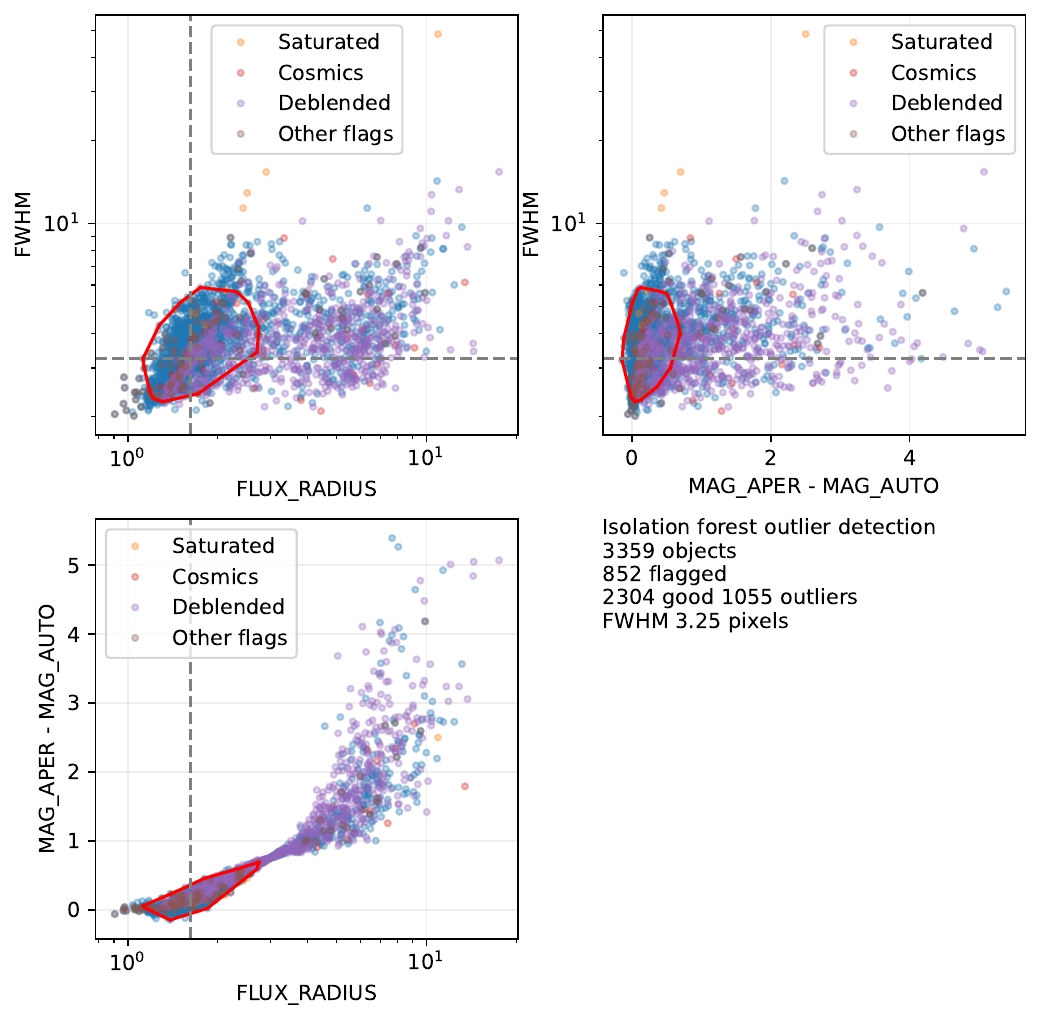}
\caption{Distribution of \software{SExtractor}-measured shape parameters for the objects detected in oversampled (left panel) and nearly critically sampled (right panel) images. Points of different colors represent various flagged objects -- the ones containing saturated pixels or cosmic ray hits inside isophotal aperture, the ones marked as deblended, and having other extraction flags set by \software{SExtractor}. Red outline marks the region selected as ``good'' by isolation forest outline rejection algorithm, as described in Section~\ref{sec:detection}.} 
\label{fig:prefilter}
\end{figure*}

While \software{photutils}\citep{photutils} does contain an extensive set of routines for object detection written in pure Python, we decided, primarily for performance reasons, to keep using \software{SExtractor}, which is highly optimized and memory efficient even for largest images, for this step in \software{STDWeb}. To run it, we construct a minimal set of configuration parameters based on the values optionally specified by the user like smoothing (``filtering'', in \software{SExtractor} terminology) kernel size, background mesh size, etc. As \software{SExtractor} does not internally support masking, we set the image values corresponding to user-specified ``custom mask'' to \texttt{NaN} to exclude them from background estimation. The objects with footprints containing the pixels masked by automatic routines (so, either saturated or marked as cosmic rays) are still properly detected, but flagged in the output so that consecutive processing may ignore them. For the visual quality control, we keep and present to the user the segmentation map, estimated background and background RMS maps, as well as filtered image produced by \software{SExtractor}. 

We also output several shape-related parameters for detected objects that are then used for simple pre-filtering of non-point-source detections in the image. After some experiments we selected three parameters produced by \software{SExtractor} that are most sensitive to shape variations and flux distribution inside the object footprint:
\begin{itemize}
    \item \texttt{FWHM\_IMAGE} -- Full Width at Half Maximum, in pixels
    \item \texttt{FLUX\_RADIUS} -- Half Flux Radius, in pixels
    \item \texttt{(MAG\_APER - MAG\_AUTO)} -- the difference between magnitudes measured in fixed circular and Kron-like elliptical apertures
\end{itemize}
The space formed by these parameters is shown in Figure~\ref{fig:prefilter} for the images with different FWHM values and for various object flags. It is clear that these parameters represent slightly different shape aspects, and thus may be used to filter out various problematic detections corresponding to artefacts or blended sources. To do so, we implemented simple outlier rejection procedure based on isolation forest anomaly detection algorithm\citep{isolationforest} as implemented in \software{scikit-learn} package\citep{scikit-learn}. Figure~\ref{fig:prefilter} shows that the outlier score produced by the algorithm nicely complement the flags set by cosmic ray detection, saturation, and \software{SExtractor} for various artefacts. Therefore we additionally flag the outliers of the algorithm so that later they may be excluded from various steps of the analysis as problematic detections. The summary of all object flags as implemented in \software{STDWeb} is given in Table~\ref{tab:flags}.

\begin{table}
\centering
\begin{tabular}{lp{5.5cm}}
\toprule
\bfseries Bitmask & \bfseries Meaning
\\ \midrule

\multicolumn{2}{l}{Flags from \software{SExtractor}}
\\ \midrule
\texttt{0x001} & Aperture flux is significantly affected by nearby stars or bad pixels
\\ 
\texttt{0x002} & Object is deblended
\\ 
\texttt{0x004} & Object is saturated
\\ 
\texttt{0x008} & Object footprint is truncated
\\ 
\texttt{0x010} & Object aperture data are incomplete
\\ 
\texttt{0x020} & Object isophotal data are incomplete
\\ 
\texttt{0x100} & Object footprint contains masked pixels
\\ 

\\ \multicolumn{2}{l}{Flags from aperture photometry}
\\ \midrule
\texttt{0x200} & Photometric aperture contains masked pixels
\\ 
\texttt{0x400} & Local background annulus does not have enough good pixels

\\ \multicolumn{2}{l}{Flags from higher-level analysis}
\\ \midrule
\texttt{0x800} & Object is classified as outlier by pre-filtering routine
\\\bottomrule
\end{tabular}
\caption{Object flags as set at various stages of the analysis.}
\label{tab:flags}
\end{table}

As \software{SExtractor} aperture photometry is somewhat limited\citep{2007PASP..119.1462B,2013PASP..125...68A}, and as it does not allow to perform forced photometry at user-specified positions, we implemented an additional photometric measurement step based on \software{photutils} routines\citep{photutils}. It also allowed us to implement proper local background estimation inside an annulus of user specified size. In order to facilitate mostly automatic use, we automatically select the aperture radius to be equal to image FWHM, which is a good compromise between signal to noise (S/N) ratio and source confusion in denser stellar fields, and place sky annulus between 5 and 7 FWHM values so that it is large enough for accurate local background estimation. These values may also be adjusted by the user, if needed. As an estimate of image FWHM, we use median value of $2\cdot$\texttt{FLUX\_RADIUS} for all unflagged objects having S/N$>$20, thus rejecting artefacts and non-point-source objects, in a way similar to implemented e.g. in \software{PSFEx}\citep{psfex}.
Using additional aperture photometry step allows us to also perform forced photometry at the positions specified by the user, e.g. corresponding to transients with already known coordinates, with exactly the same parameters like aperture and background subtraction methods, thus ensuring common effective zero point for all measurements.

After object measurement, we exclude from the object list the detections with sufficiently large photometric errors, thus keeping only the ones that have at least user-specified signal to noise ratio (by default 5, corresponding to magnitude errors smaller that 0.2). Together with rejection of detections having too small footprints (\texttt{DETECT\_MINAREA} parameter of \software{SExtractor}) and pre-filtering routine described above it allows to reject obvious pixel-level and extended artefacts, as well as to keep only objects with properly measured flux.

\subsection{Astrometric calibration}
\label{sec:astrometry}

If the FITS header does not contain usable astrometric solution, or if the user suspects it to be unreliable, the software initiates blind solving for it using local instance of \software{Astrometry.Net}\citep{astrometrynet} code with 2MASS indices that are sufficient for most of deeper images. In order to ensure that astrometric solution uses the same positions as other steps of our analysis, as well as for optimizing the performance, we run the code directly on the list of object positions detected on previous step, additionally filtering out all flagged objects to avoid spurious matches due to non-stellar objects and artefacts. Optionally the user may interactively provide the limits on the sky position and range of pixel scales used for finding the solution, otherwise the full search for it is performed. We also run the code twice, in order to properly refine the solution to produce WCS with second-order SIP order distortions included.

As soon as preliminary astrometric solution is known, either from blind matching or from original FITS header, we request the catalogue from Vizier that will be used for photometric calibration (see Section~\ref{sec:photometric} for more details) and perform additional astrometric refinement using \software{SCAMP}\citep{scamp} code by passing both the catalogue and list of unflagged detected object positions to it. This way we ensure that the astrometric solution is always correct down to accuracy required for photometric calibration and image subtraction steps.

\subsection{Reference catalogues and photometric calibration}
\label{sec:photometric}

\begin{figure*}
\centering
\includegraphics[width=0.49\linewidth]{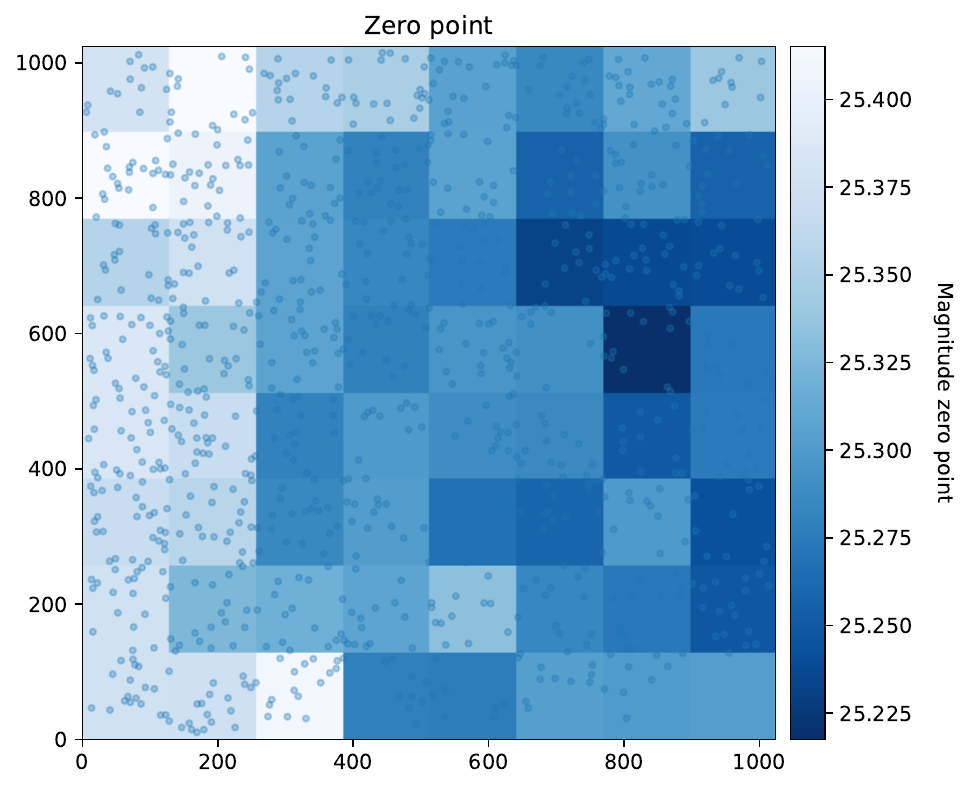}
\includegraphics[width=0.49\linewidth]{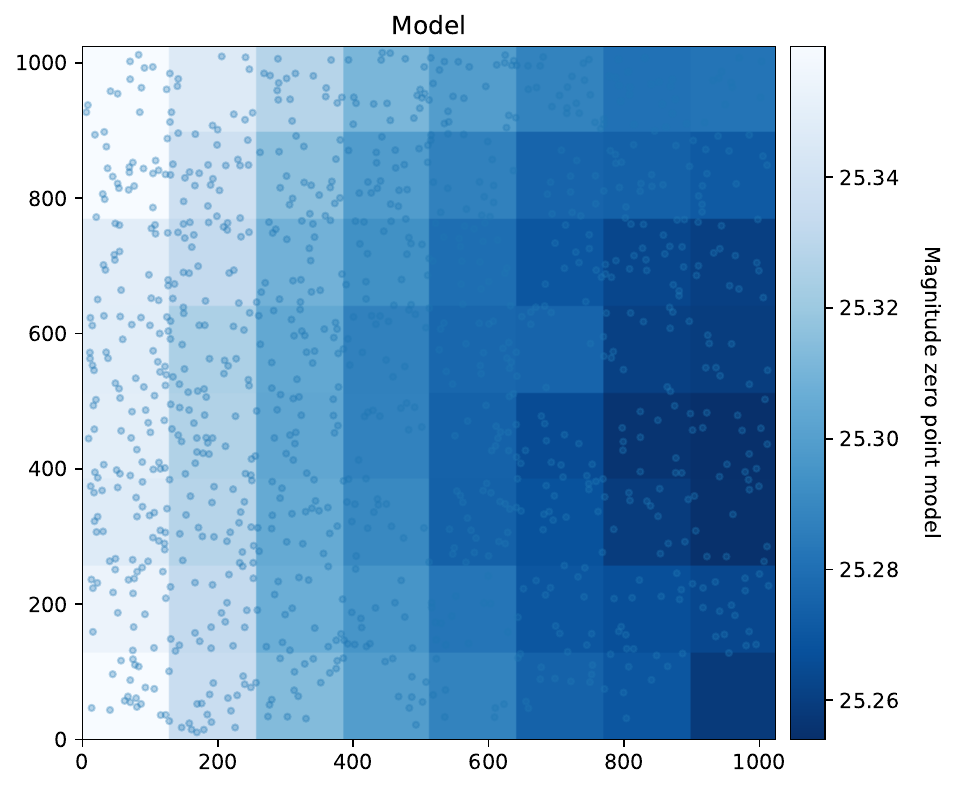}
\includegraphics[width=0.49\linewidth]{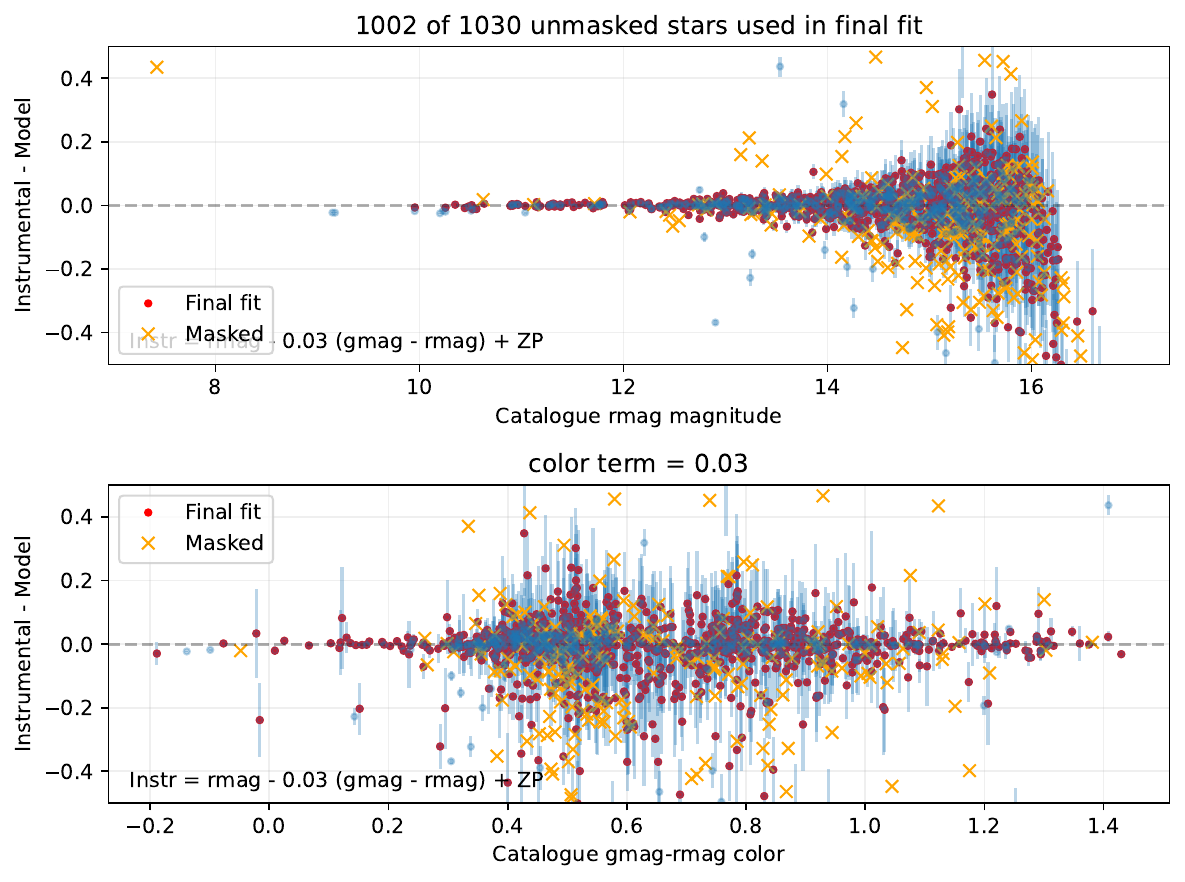}
\includegraphics[width=0.49\linewidth]{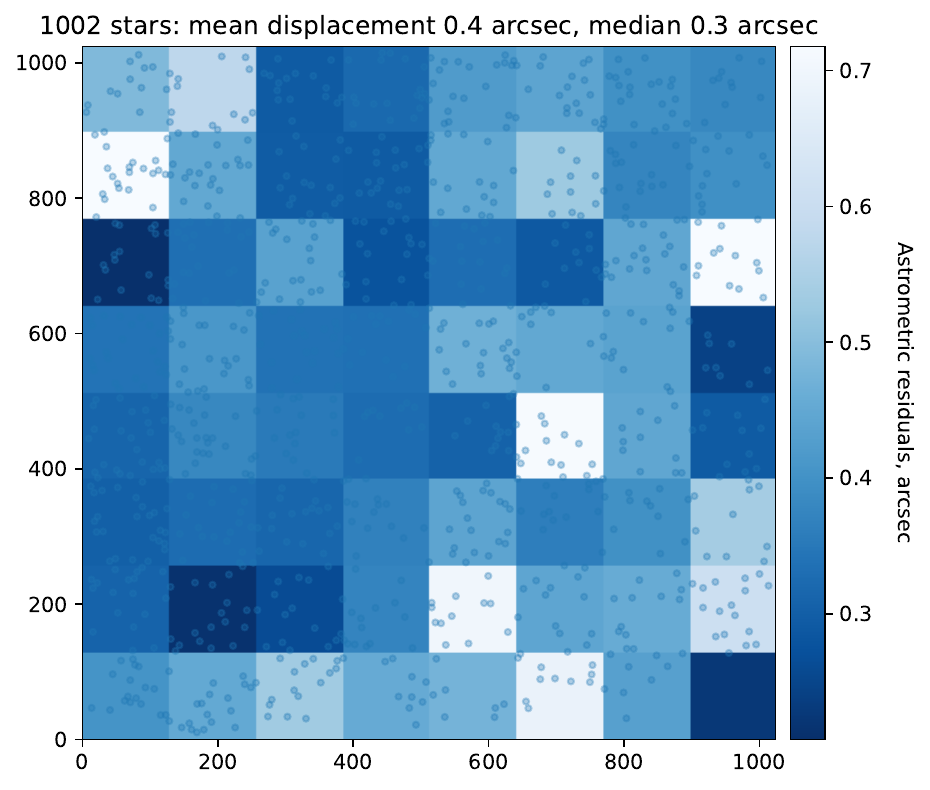}
\caption{Example of some diagnostic plots produced by \software{STDWeb} while fitting photometric model using both positionally-dependent zero point and color term. The panels show the original map of differences between instrumental and catalogue magnitudes (upper left), the model built for it using second order spatial polynomial and linear color term (upper right), residuals between the data and the model as function of catalogue magnitude and catalogue $g-r$ color index (lower left), and the map of astrometric displacement between matched objects and catalogue stars (lower right).} 
\label{fig:photomodel}
\end{figure*}

\begin{figure*}
\centering
\includegraphics[width=0.49\linewidth]{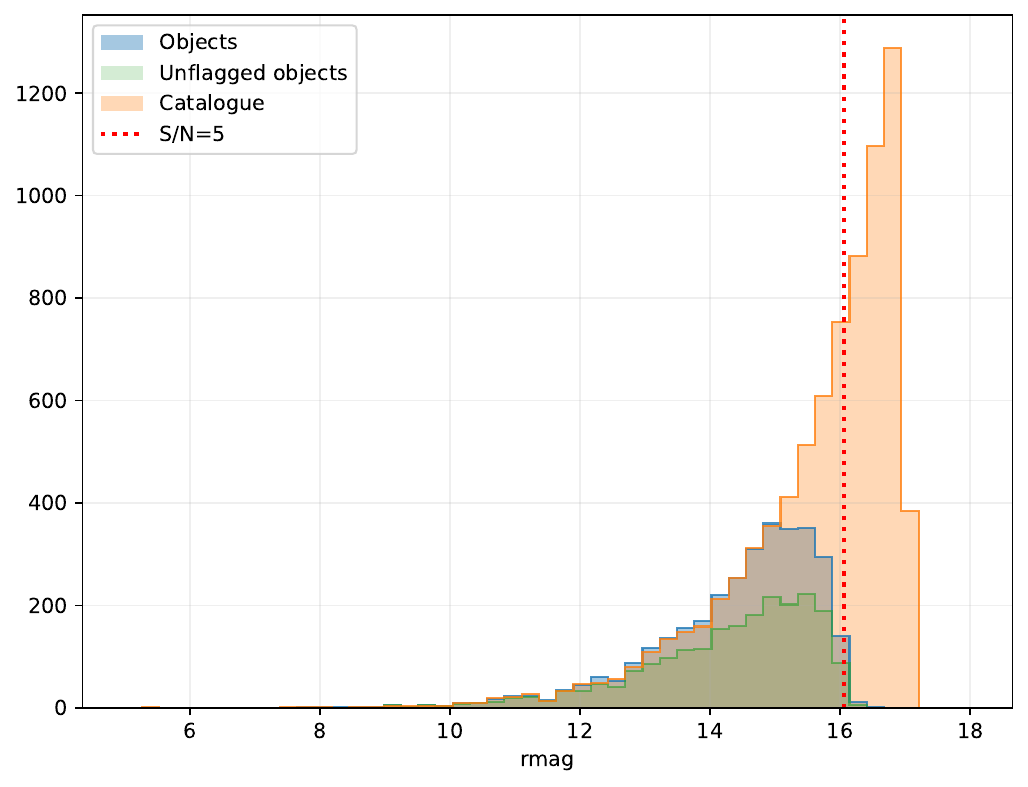}
\includegraphics[width=0.49\linewidth]{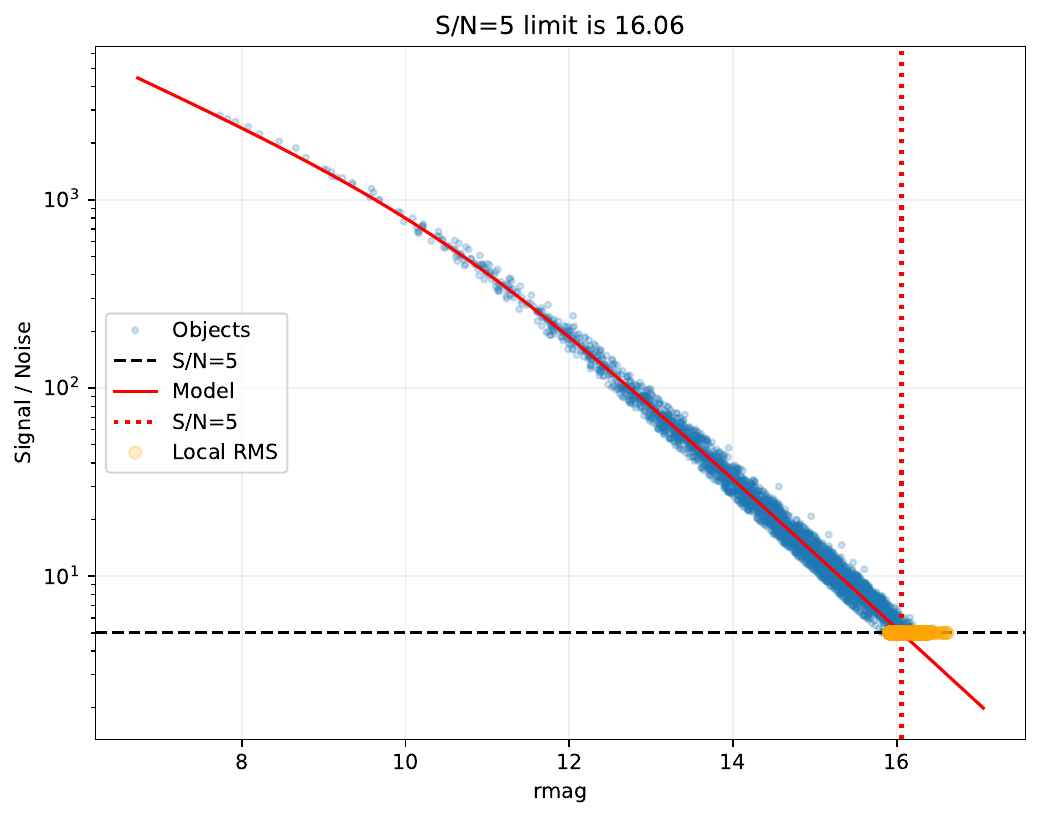}
\caption{Example of diagnostic plots produced by \software{STDWeb} for the determination of detection limit. Left panel -- the histograms of calibrated magnitudes for detected objects (both total and unflagged ones only), and catalogue magnitudes for stars inside the image. Right panel -- signal to noise ratio vs calibrated magnitude plot of detected objects, fitted with the noise model. Detection limit is defined as an intersection of the model curve with selected level (S/N=5 in this example). Also plotted (orange dots) are detection limits estimated for individual objects solely from positionally-dependent background flux errors inside the aperture.} 
\label{fig:limit}
\end{figure*}

\begin{figure}
\centering
\includegraphics[width=1.0\linewidth]{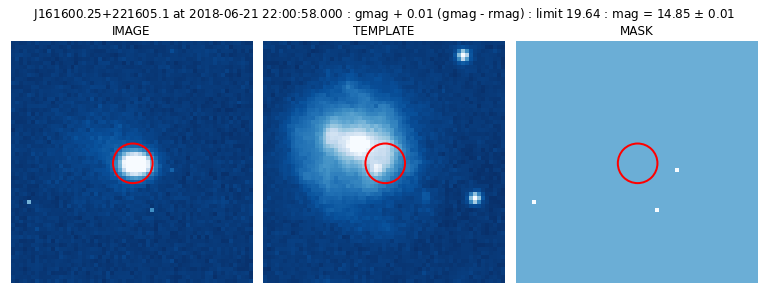}
\caption{Diagnostic plot for the forced photometry at user-specified position, containing three cutouts centered on it -- original image (left panel), atlas image for visual comparison (either Pan-STARRS or SkyMapper images from \texttt{HiPS2FITS}\cite{hips2fits} service are used, depending on the hemisphere) (middle panel), and mask image that marks the positions of masked pixels (right panel). Reticle (red circle) marks the position of the target in all panels. Title summarizes the relevant information like calibrated magnitude of the source, effective passband, local detection limit, timestamp, etc.}
\label{fig:target_direct}
\end{figure}

In order to perform photometric calibration, we use one of larger catalogues covering large portions of the sky and available in CDS Vizier\citep{vizier} -- Pan-STARRS DR1\citep{ps1surveys}, SkyMapper Southern Survey DR4\citep{skymapper}, Gaia eDR3\citep{gaiaedr3}, compilative all-sky ATLAS-REFCAT2\citep{refcat2} catalogue, and the catalogue of synthetic photometry computed from Gaia low-resolution spectra (Gaia DR3 Syntphot\citep{syntphot}). The catalogues are requested from Vizier on the fly using \software{astroquery}\citep{astroquery} package, optionally limiting to the range of magnitudes relevant for the image to restrict download size. In order to ensure common photometric system for calibrations done using different catalogues, we derived a set of transformation equations\footnote{Conversion coefficients are available directly in the code at GitHub at \url{https://github.com/karpov-sv/stdpipe/blob/master/stdpipe/catalogs.py}} between their individual passbands, Pan-STARRS and Johnson-Cousins systems using large set of Landolt standards assembled in \citet{Pancino_2022}. When appropriate, these transformations were derived using two base colors (e.g. $g-r$ and $r-i$ for Pan-STARRS and SkyMapper magnitudes conversion to Johnson-Cousins system) to ensure the transformations are valid for both hot and cold stars. On the other hand, for converting between closer systems (e.g. Sloan or SkyMapper to Pan-STARRS) single color regression is usually enough.
This way, we have a set of catalogs covering all sky and sufficiently complete for both brighter (Gaia DR3 Syntphot) and fainter (Pan-STARRS and SkyMapper) stars, providing measurements in both Pan-STARRS and Johnson-Cousins systems all calibrated to common zero points. While technically it is also possible to derive a set of approximate transformations to and from Gaia eDR3 set of filters, we decided not to do it, leaving it as a separate option for calibrating unfiltered or very wide passband images, using $G$ magnitude and $BP-RP$ color as a basis.

We perform photometric calibration of the image by building the global zero point model applicable to all detected objects. To do so, we have to take into account various effects like a difference between instrumental and catalogue passbands, positional dependence of aperture correction, atmospheric extinction and uncorrected detector sensitivity, etc. To do so, we use the following photometric model:
\begin{equation}
    \textnormal m_{\textnormal{calib}} = \textnormal m_{\textnormal{instr}} + \textnormal{ZP}(x, y) 
    + C \cdot \textnormal{Color}
    \label{eq:std-calib}
\end{equation}
where $\textnormal m_{\textnormal{calib}}$ is the magnitude of the object in the catalogue system, $m_{\textnormal{instr}} = -2.5\cdot\log_{10}{(\textnormal{ADU})}$ is the instrumental magnitude computed from the measured object flux in the image, $\textnormal{ZP}(x, y)$ is positionally dependent zero-point, $C$ is the color term and $\textnormal{Color}$ is the color of the object in the catalogue\footnote{While \software{STDPipe} photometric solver also supports fitting positionally dependent additive flux term to account for e.g. biased background estimation, we decided to avoid using it in \software{STDWeb} for reducing the number of free parameters which is critical for typical narrow-field images with not so many stars present.}.  
Positionally-dependent zero point is modelled as a spatial polynomial of an user-specified order in pixel coordinates, while the choice of catalogue color depends on the desired passband used for a primary catalogue magnitude. So, for primary magnitudes in Pan-STARRS $grizy$ system the $g-r$ color is used (except for $z$ filter where we selected $r-i$ color as more appropriate), while for Johnson-Cousins primary magnitudes the color is $B-V$, and $BP-RP$ is the one for Gaia $G$ primary filter. The choice of the primary filter depends on the filter used to acquire the image, and may either be set automatically based on FITS header keywords, or directly specified by the user alongside with the choice of appropriate reference catalogue. Optionally, the software also performs color term diagnostics by consecutively using all supported filters for a given catalogue as a primary, and computing corresponding color term. Then the filter that minimizes the color term -- thus, corresponding to the passband closest to the instrumental system of the image -- may be selected for the calibration.

The model is fit using robust fitter with iteratively rescaled errors to account for potentially underestimated uncertainties of the catalogue and measurements, with additional intrinsic scatter of measurements up to 0.01 mag also being fit as a free model parameter to account for e.g. uncorrected fixed pattern errors. The fitting involves all pairs between detected objects and catalogue stars positionally matched within 0.5$\cdot$FWHM from each other. No additional criteria is applied to ensure the uniqueness of the matching, as non-unique matches just introduce the outliers in the fit that will be iteratively rejected. Instead, we pre-process the catalogue in order to exclude close pairs of stars (closer than 2$\cdot$FWHM to each other) that would potentially correspond to blended objects in the image, to avoid using them in the fit. Moreover, we exclude from the fit all flagged objects corresponding to problematic, saturated, or pre-filtered detections in order to further reduce the amount of outliers. The fitting produces an extensive set of diagnostic plots, shown in Figure~\ref{fig:photomodel} that may be used for checking the fit quality.

Once constructed, this photometric model is used to calibrate the measurements of all detected sources, as well as forced photometry at specified target positions (see Figure~\ref{fig:target_direct}).

\subsection{Detection limit}
\label{sec:limit}

There are several possible approaches for defining the detection limit -- the brightness of the faintest detectable source -- of the image, from synthetic source injection to the analysis of noise within photometric aperture. In 
\software{STDWeb} we adopted the based on the studying the dependence of signal to noise ratio on magnitude in a list of objects detected and measured in the image, as shown in right panel of Figure~\ref{fig:limit}. To derive it, the software fits the set of corresponding S/N values for all objects with the parametric model consisting of constant background noise and Poissonian noise of the object flux that is proportional to the flux square root. While the model is essentially the same as used by photometric routines for estimating the measurement error, the plot still has some scatter due to positional dependence of the zero point. Thus, the estimated global detection limit corresponds to some mean value over the whole image, while local values may slightly deviate from that. The code also tries to estimate the limits for individual objects solely from the (positionally dependent) noise inside the aperture multiplied by a given S/N ratio (thus, corresponding to the flux that would be detectable with that significance above the background noise), but this approach slightly overestimates the limit due to not taking into account noise contribution from the object flux itself, as seen in right panel of Figure~\ref{fig:limit} (orange dots).

The histogram of magnitudes of detected objects in comparison with the ones for catalogue stars inside the image may also be used for both estimating the detection limit, as well as completeness and purity of the detections, as shown in left panel of Figure~\ref{fig:limit}. In this specific example, which corresponds to the same image as right panel of Figure~\ref{fig:prefilter}, most of flagged objects represent the blended ones that are marked as outliers by pre-filtering routine.

\section{Transient detection}
\label{sec:transients}

Major efforts in creation of \software{STDWeb} have been put in implementation of transient detection routines, as they are the primary motivation behind the original \software{STDPipe}.
The code implements two different approaches for detecting the transients in the image -- the one based solely on comparing the lists of detected objects with different catalogues, and the one involving subtraction of reference image in order to find pixel-level differences between them. Details of their implementations are given below.

\subsection{Simple catalogue-based transient detection}
\label{sec:transients_simple}

\begin{figure}
\centering
\includegraphics[width=1.0\linewidth]{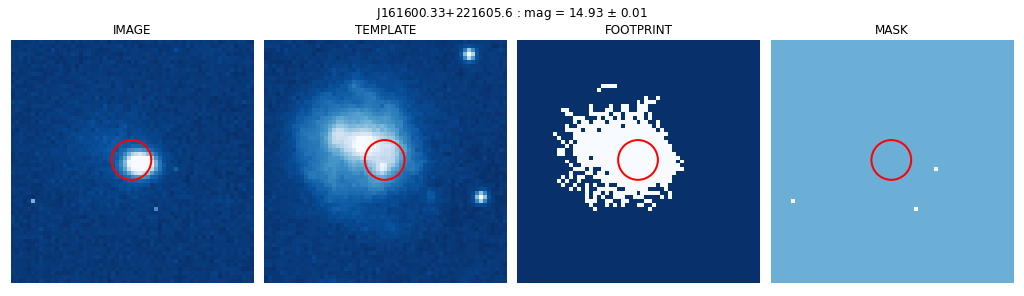}
\caption{Diagnostic plot for the candidate selected by simple transient detection routine, containing four cutouts centered on its position -- original image (first panel), atlas image for visual comparison (either Pan-STARRS or SkyMapper images from \texttt{HiPS2FITS}\cite{hips2fits} service are used, depending on the hemisphere) (second panel), the detection footprint from \software{SExtractor} segmentation map (third panel), and mask image that marks the positions of masked pixels (last panel). Reticle (red circle) marks the position of the target in all panels. Title summarizes the relevant information like calibrated magnitude of the source, effective passband, local detection limit, timestamp, etc. In this example, the code properly detects the transient, but due to it being superimposed on top of extended galaxy, the candidate centroid position is biased, and the isophotal footprint includes the flux from the galaxy itself, thus affecting the object' shape parameters that may lead to it being rejected by shape-based classifier as described in Section~\ref{sec:detection}.}
\label{fig:transient_simple}
\end{figure}

For detecting brighter transients that are not situated on a complex background or in a dense stellar fields, the approach based on direct comparison of objects detected in the image with the lists of known catalogue stars may be sufficient. It is implemented in \software{STDWeb} as a ``Simple transient detection'' section of task analysis, and involves the following steps.

The analysis is based on the list of objects detected on the original object detection step, that includes only the ones that have aperture flux measured with signal to noise ratio better than user-specified threshold. They all have astrometrically calibrated positions and magnitudes calibrated to standard (Pan-STARRS or Johnson-Cousins) system, with instrumental photometric system characterized by linear color term.
First, the code rejects all the objects that have any flag set (see Table~\ref{tab:flags}) except for \texttt{0x002} (deblended) or \texttt{0x100} (isophotal footprint contains masked pixels), thus rejecting all saturated ones, the ones marked as cosmic rays, and the ones rejected by shape-based classifier (see Section~\ref{sec:detection}). The latter criterion may be optionally relaxed,  in order to be able to e.g. detect the transients that are in close proximity of other objects, or just in the images where shape-based classifier rejects too many good objects.

Then, the user is able to restrict the search to just a portion of the sky by specifying the center and radius corresponding e.g. to known error box where the transient is expected. If so, the filtering with these parameters is performed, and the objects falling outside the specified cone are excluded from the following analysis. Next, simple ``multi-image mode'' may also be enabled -- if several images covering the same sky position at approximately the same time are uploaded to \software{STDWeb}, then the user may specify request to reject any object that is not apparent in all these images simultaneously. This is done by specifying task IDs of corresponding additional images that should be already photometrically calibrated, performing the positional cross-match on them, and rejecting the objects that have no counterparts in all of them.

Next, the remaining list of objects is cross-matched with a list of catalogues (Gaia eDR3, then Pan-STARRS DR1, then SkyMapper DR4) using CDS XMatch\citep{xmatch} service within 2$\cdot$FWHM radius. As the list of catalogue fields returned by it is a bit limited compared to regular CDS Vizier\citep{vizier}, we do not perform proper augmentation of cross-match results with Pan-STARRS or Johnson-Cousins magnitudes, but instead just select the closest catalogue passband, apply AB to Vega zero point difference if necessary, and compare the brightness of detected source with it. We reject all the matches where the object is not brighter that catalogue star by a specified limit (2 magnitudes by default). This way, we reject all positional matches except for the ones where catalogue star is too faint compared to detected transient candidate in order to both avoid spurious matches with faint background catalogue sources, and keep the transients related to significant increase of the flux from known objects (e.g. nuclear transients or stellar flares). As a final step, if the timestamp of the image is known, remaining transient candidates are checked against the positions of known Solar System objects at that moment using IMCCE SkyBoT\citep{skybot} service, and the candidates positionally matched within 10$''$ are rejected. No brightness comparison is performed between the candidate and Solar System object on this step. Instead, this checking may be completely disabled by the user.

The results of the search are presented to the user as a set of candidates, each accompanied by the cutouts from original image centered on it, reference image from relevant sky survey (either Pan-STARRS or SkyMapper) in the closest passband, the detected object footprint, and the mask image, as shown in Figure~\ref{fig:transient_simple}. The whole list is also directly downloadable in a tabular format.

\subsection{Image subtraction}
\label{sec:subtraction}

\begin{figure*}
\centering
\includegraphics[width=1.0\linewidth]{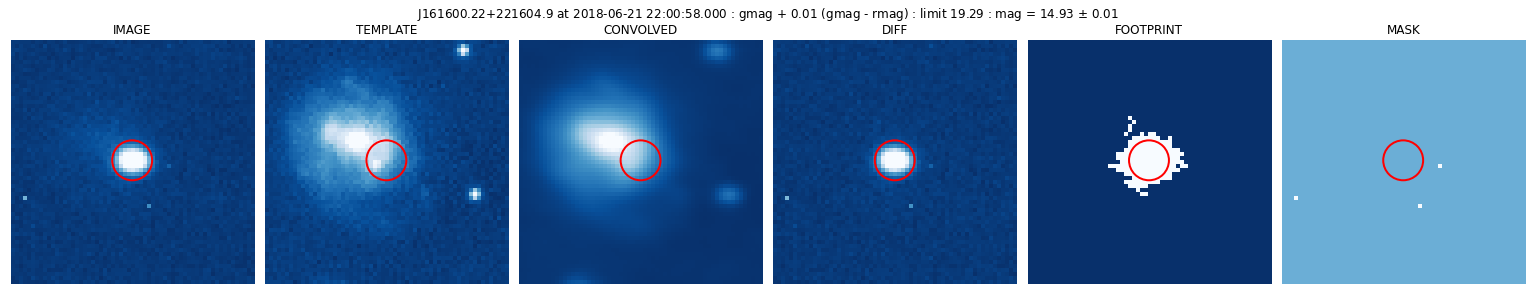}
\caption{Diagnostic plot for the candidate selected by image subtraction based transient detection routine, containing six cutouts centered on its position -- original image (first panel), template image (second panel), template image convolved with the kernel to match the resolution and flux scale of original image (third panel), difference image (fourth panel)
the detection footprint from \software{SExtractor} segmentation map (fifth panel), and mask image that marks the positions of masked pixels (last panel). Reticle (red circle) marks the position of the target in all panels. Title summarizes the relevant information like calibrated magnitude of the source, effective passband, local detection limit, timestamp, etc. 
The candidate shown here is the same as in Figure~\ref{fig:transient_simple} -- it is clearly seen that this time its position is unbiased and the footprint does not include extra regions, as the underlying galaxy was properly subtracted.}
\label{fig:transient_subtracted}
\end{figure*}

\begin{figure}
\centering
\includegraphics[width=1.0\linewidth]{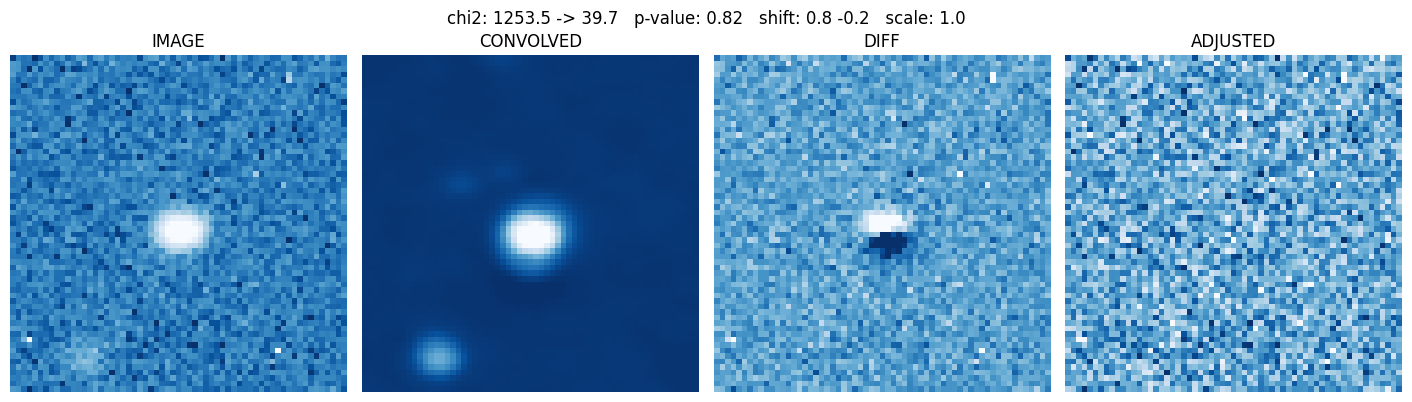}
\includegraphics[width=1.0\linewidth]{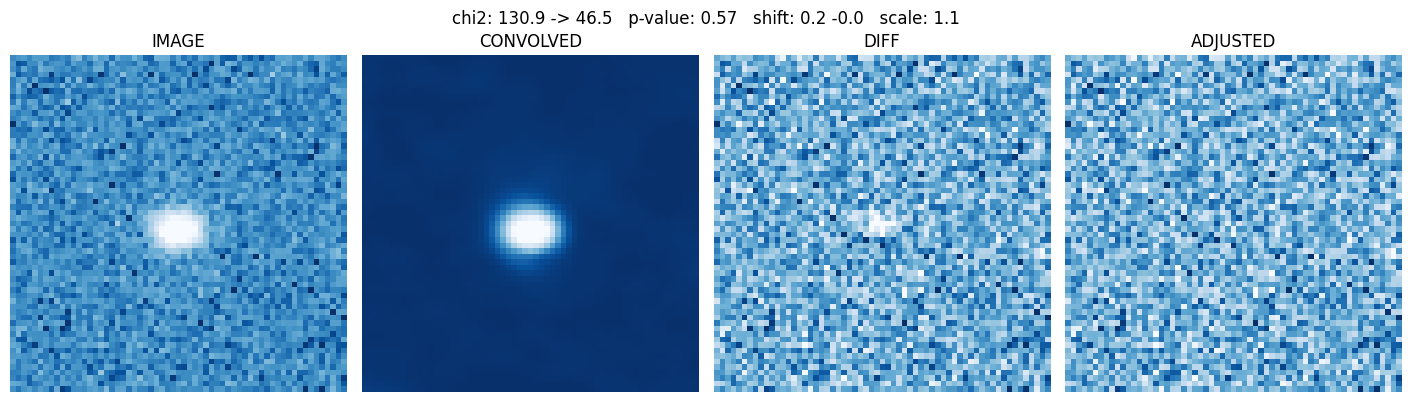}
\includegraphics[width=1.0\linewidth]{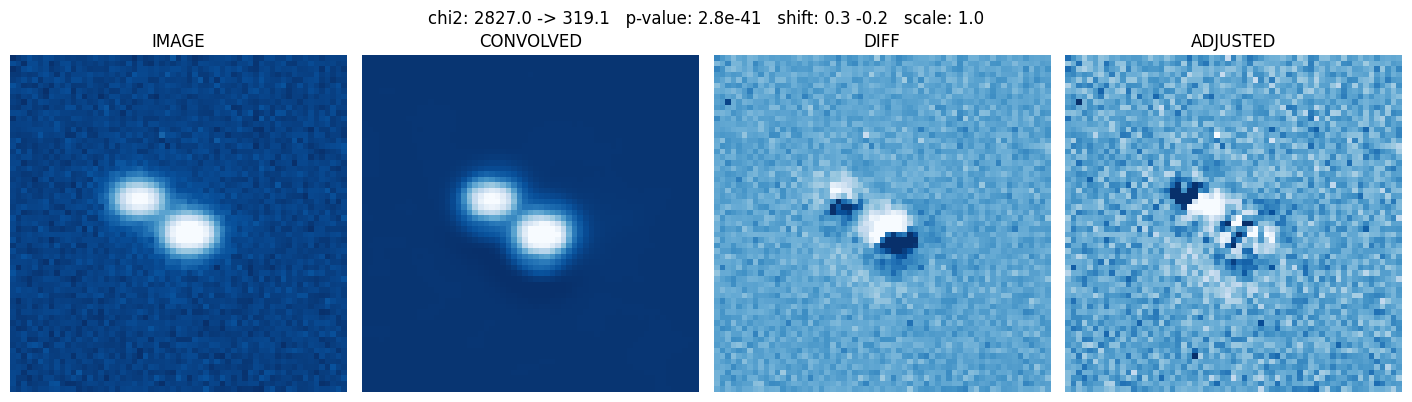}
\caption{Examples of successful cutout-level sub-pixel adjustment routine that is able to fully eliminate the ``dipole'' artefact related to position misalignment between the images (upper row) and the candidate related to slight change of the object brightness (middle row), and mostly eliminate the candidate in a close pair where relative position of the stars changed due to proper motion (so, after the adjustment, second component ``dipole'' became even move prominent). Every row shows the cutouts from original image, template convolved with matching kernel, original difference and adjusted difference images. The adjustment is performed within 2$\cdot$FWHM x 2$\cdot$FWHM central part of the cutout only. The row title shows the values of $chi^2$ in this region before and after the adjustment, determined positional shift and flux scaling.}
\label{fig:adjust}
\end{figure}

Image differencing, or reference image subtraction, is a now \textit{de facto} standard way of detecting transients in astronomical images. There are two primary approaches for that being used in modern large scale sky surveys and their pipelines -- the one based on the determination of optimal convolution kernel that matches the PSF of one image with the other directly in pixel space\citep{alard_lupton_1998} (often called Alard-Lupton method), and the one that instead derives the optimal point source detection statistics\citep{zogy} (Zackay-Ofek-Gal-Yam, or ZOGY, method). In \software{STDWeb}, we support both of these methods, with the primary one being the former, as implemented in publicly available \software{HOTPANTS}\citep{hotpants} code, due to its ability to work without prior knowledge of exact PSFs and relative flux scale of the images.

Image subtraction requires a reference, or template, image with comparable or preferably better detection limit. Modern large-scale sky survey experiments produce such templates as part of their routine observations, by stacking series of past images covering the same sky position, but for smaller telescopes that perform just occasional follow-up observations this approach is not available. Therefore, transient detection on such images require either acquiring the reference image by observing the same location after the transient faded down, which is impractical in most cases, or falling back to publicly available archival images from some of the large scale sky surveys that produced sky atlases, like Pan-STARRS 3Pi survey\citep{ps1surveys}, or DESI Legacy Surveys\citep{legacysurveys}, with obvious limitations being different epoch of observations, different passbands, and sometimes non-trivial imaging artefacts in these data. \software{STDWeb} implements acquiring template images from both of these data sources, automatically performing their re-projection and mosaicking, as well as handling some quirks of their data formats (like non-linear scaling of Pan-STARRS images) and producing corresponding masks. However, as neither of these two surveys cover the whole Southern sky, as an optimistic fallback data source \software{STDWeb} also supports getting the data from several surveys like Dark Energy Survey, SkyMapper or ZTF,  available through HiPS2FITS\citep{hips2fits} service, the same that powers e.g. Aladin Sky Atlas\citep{aladin}. The service conveniently allows getting FITS images already projected onto requested WCS pixel grid, but lacks any means of acquiring corresponding mask information, thus limiting the usability of this approach.
Also, \software{STDWeb} allows user to directly upload custom template image as a separate FITS file that will be automatically aligned with original image for the subtraction.

Upon acquisition and re-projection of template image, the code tries to automatically select the best set of parameters for \software{HOTPANTS} based on image FWHM (it is estimated as part of object detection step, see Section~\ref{sec:detection}), saturation level, similar to how it is done in \citet{iptf}.
Moreover, it constructs the noise model for both image and template
by combining their estimated background RMS with corresponding gain values, thus allowing the code to work properly on background-subtracted images (and, to some extent, even on the images with unknown or incorrect gain value). It uses masks from image and template to exclude bad image regions from the fitting, and directly passes to the code the positions of good objects to be used for selecting best positions for placing the sub-stamps used to derive the convolution kernel. We also set up \software{HOTPANTS} to always convolve the template to match the image, even if it is not optimal, in order to keep the same zero point for aperture photometry on difference image. However, the user may directly modify all input parameters used to call the code, and so revert this behaviour, as well as adjust any other aspect of \software{HOTPANTS} operation.

In order to reduce the impact of positionally-dependent PSF shape or improper flat-fielding, the software automatically splits the image into a set of chunks of approximately specified size with some overlaps to avoid any position inside it being too close to the chunk edge, processes these chunks separately, and then combines the results together. Specifically, chunk processing includes getting the difference image and noise model for it from \software{HOTPANTS} output, performing noise-weighted object detection in difference image using \software{SExtractor}, applying the same pre-filtering of detections as described in Section~\ref{sec:detection} using the classifier derived from original image, and performing the aperture photometry at candidate positions using the same aperture settings as used in original image in order to keep the same zero point. This way, we derive the list of transient candidates that may be further filtered, like in Section~\ref{sec:transients_simple}, using positional criteria, or cross-matching with Vizier catalogues or Solar System object positions (both disabled by default, as most of stationary objects are subtracted and not being detected as candidates). The final list of candidates is directly available for downloading in tabular format, and is presented to the user like shown in Figure~\ref{fig:transient_subtracted}.

Image subtraction is not perfect, and prone to various kinds of artefacts related to both numerical instabilities of kernel fitting routine (typically producing ring-shaped artefacts) and slight differences of object positions between the images, e.g. due to astrometric misalignments or proper motions of the stars (typically producing characteristic ``dipole'' artefacts, consisting of overlapped positive and negative components of similar amplitude). While both of them may be, in most cases, successfully filtered out by a shape-based classification (see Section~\ref{sec:detection}), we have an additional routine implemented to specifically target the latter ``dipole'' artefacts. The routine is based on 
sub-pixel adjustment of the relative shift, along with minor adjustment of the flux scale, between convolved template and original image in order to minimize the difference between them
in the region around the candidate.
If the routine (limited to maximum of 1 pixel shift, and no more than 30\% flux scale adjustment) is able to reduce the $\chi^2$ of the difference by more than three times, or if the final $p$-value is exceeding $0.01$, as shown in Figure~\ref{fig:adjust}, then the candidate is considered spurious and is also rejected. Allowing minor optimization of flux scale hinders the ability to detect low-amplitude variability of the objects, but we consider it a good trade-off with overcoming the artefacts due to passband differences between the image and the template.
While this algorithm is by no means a proper replacement for machine-learning artefact rejection  methods like convolutional neural networks\citep{ztf_rb,otrain}, it is generic enough to be used on arbitrary images without any preliminary training. 

If the position of the transient is known in advance, \software{STDWeb} also allows skipping transient detection part of image subtraction, and directly performing the forced photometry at its position in difference image.

\section{Conclusions and future development}
\label{sec:conclusions}

We present \software{STDWeb}, a simple web-based photometric calibration and transient detection tool that allows semi-automated and interactive analysis of arbitrary sky images in order to perform their astrometric and photometric calibration, forced photometry, and transient detection, presenting the user with rich set of controls and diagnostic outputs. It is intended to be used both as a simple user-level tool, and as a part of data infrastructure of projects involving diverse sets of images from various sources, like telescope networks. It is already deployed as a standard image processing tool within GRANDMA network \citep{grandmao41,grandma_ztf,grandmao42}.

The niche occupied by \software{STDWeb} lies between the diverse zoo of lower-level utilities and libraries like \software{SExtractor}\citep{sextractor}, \software{DAOPHOT}\citep{daophot} or \software{Photutils}\citep{photutils}, and powerful but complicated integrated environments like \software{IRAF}\citep{iraf}, and is close to more specialized batch processing pipelines like \software{AutoPhOT}\citep{autophot}, \software{PHOTOMETRYPIPELINE}\citep{photometrypipeline}, or \software{VaST}\citep{vast} that aim to solve some specific task in an automated manner. However, due to its flexibility and interactivity \software{STDWeb} is also similar to much more user-friendly GUI based packages like \software{AstroImageJ}\citep{astroimagej}, \software{Aperture Photometry Tool}\citep{apt}, \software{PixInsight}\footnote{\url{https://pixinsight.com}}, or \software{SIPS}\footnote{\url{https://www.gxccd.com/cat?id=146&lang=405}}. In contrast to them, \software{STDWeb} does not have advanced image stacking or processing capabilities and does not support image sequence analysis that is critical for variable stars photometry (typical application field of these packages), but, on the other hand, it is a web application focused more on transient detection and accurate ensemble photometry of a single image that does not require local installation and is accessible with any device. 
Moreover, \software{STDWeb} is open source and may be easily deployed and adapted for specific tasks and project environments, in contrast to e.g. proprietary
\software{AAVSO VPhot}\footnote{\url{https://apps.aavso.org/vphot/}} web tool.

While already feature complete, the code of \software{STDWeb} continues being actively developed. Directions for future improvements include overcoming some of its shortcomings listed above like better cross-image analysis that is important both for checking whether the detected transient candidates  are real, and quick examination of their temporal behaviour, better rejection of image artefacts, as well as some of the advanced image pre-processing algorithms for removing low-frequency contaminations like fringes or nebulous background components. Another direction is the inclusion of PSF photometry when it is appropriate, as well as overall optimization of the code so that it is usable even on largest images with many undersampled stars, like the ones produced by wide-field sky survey telescopes.
The user interface is also subject for future improvements, especially in the part related to quick visual checking of the transient candidates, by e.g. allowing rapid access to sky atlas images and known catalogue objects around them, lists of nearby transients from Transients Name Server\citep{tns} and alert brokers like Fink\citep{fink}, etc.

The code of \software{STDWeb} is publicly available, along with detailed installation and deployment instructions, on GitHub at \url{https://github.com/karpov-sv/stdweb}.

\begin{acknowledgements}
This work was co-funded by the EU and supported by the Czech Ministry of Education, Youth and Sports through the project  CZ.02.01.01/00/22\_008/0004596 (SENDISO).
This research made use of Astropy, a community-developed core Python package for Astronomy\citep{astropy}. This research made use of Photutils, an Astropy package for
detection and photometry of astronomical sources\citep{photutils}. This research made use of data provided by Astrometry.net\citep{astrometrynet}. This research made use of hips2fits\citep{hips2fits}, a service provided by CDS.
\end{acknowledgements}

\bibliographystyle{actapoly}
\bibliography{stdweb} 

\begin{thebibliography}{10}
\providecommand{\url}[1]{\texttt{#1}}
\providecommand{\urlprefix}{URL }
\providecommand{\eprint}[2][]{\url{#2}}
\makeatletter
\def\bibdoi{\begingroup\def\do##1{\catcode `##112\relax}\do$\do\\\do\_\do\%\do\^\expandafter\endgroup\@bibdoi}
\def\@bibdoi#1{\eprint{https://doi.org/#1}}
\makeatother

\bibitem{iptf}
Y.~{Cao}, P.~E. {Nugent}, M.~M. {Kasliwal}.
\newblock {Intermediate Palomar Transient Factory: Realtime Image Subtraction Pipeline}.
\newblock \textit{\pasp} \textbf{128}(969):114502, 2016.
\eprint{1608.01006} \bibdoi{10.1088/1538-3873/128/969/114502}
\bibitem{ztf}
E.~C. {Bellm}, S.~R. {Kulkarni}, M.~J. {Graham}, et~al.
\newblock {The Zwicky Transient Facility: System Overview, Performance, and First Results}.
\newblock \textit{\pasp} \textbf{131}(995):018002, 2019.
\eprint{1902.01932} \bibdoi{10.1088/1538-3873/aaecbe}
\bibitem{ps1pipeline}
E.~A. {Magnier}, W.~E. {Sweeney}, K.~C. {Chambers}, et~al.
\newblock {Pan-STARRS Pixel Analysis: Source Detection and Characterization}.
\newblock \textit{\apjs} \textbf{251}(1):5, 2020.
\eprint{1612.05244} \bibdoi{10.3847/1538-4365/abb82c}
\bibitem{goto}
D.~{Steeghs}, D.~K. {Galloway}, K.~{Ackley}, et~al.
\newblock {The Gravitational-wave Optical Transient Observer (GOTO): prototype performance and prospects for transient science}.
\newblock \textit{\mnras} \textbf{511}(2):2405--2422, 2022.
\eprint{2110.05539} \bibdoi{10.1093/mnras/stac013}
\bibitem{grandmao41}
S.~{Agayeva}, V.~{Aivazyan}, S.~{Alishov}, et~al.
\newblock {The GRANDMA network in preparation for the fourth gravitational-wave observing run}.
\newblock In D.~S. {Adler}, R.~L. {Seaman}, C.~R. {Benn} (eds.), \textit{Observatory Operations: Strategies, Processes, and Systems IX}, vol. 12186 of \textit{Society of Photo-Optical Instrumentation Engineers (SPIE) Conference Series}, p. 121861H. 2022.
\eprint{2207.10178} \bibdoi{10.1117/12.2630240}
\bibitem{grandma_ztf}
V.~{Aivazyan}, M.~{Almualla}, S.~{Antier}, et~al.
\newblock {GRANDMA observations of ZTF/Fink transients during summer 2021}.
\newblock \textit{\mnras} \textbf{515}(4):6007--6022, 2022.
\eprint{2202.09766} \bibdoi{10.1093/mnras/stac2054}
\bibitem{grandmao42}
I.~{Tosta e Melo}, J.~G. {Ducoin}, Z.~{Vidadi}, et~al.
\newblock {Ready for O4 II: GRANDMA observations of Swift GRBs over eight weeks in spring 2022}.
\newblock \textit{\aap} \textbf{682}:A141, 2024.
\eprint{2310.17287} \bibdoi{10.1051/0004-6361/202347938}
\bibitem{kncatcher}
D.~{Turpin}.
\newblock {Kilonova-catcher: a new citizen science project to explore the multi-messenger transient sky}.
\newblock In J.~{Richard}, A.~{Siebert}, E.~{Lagadec}, et~al. (eds.), \textit{SF2A-2022: Proceedings of the Annual meeting of the French Society of Astronomy and Astrophysics}, pp. 153--157. 2022.

\bibitem{iraf}
D.~{Tody}.
\newblock {The IRAF Data Reduction and Analysis System}.
\newblock In D.~L. {Crawford} (ed.), \textit{Instrumentation in astronomy VI}, vol. 627 of \textit{Society of Photo-Optical Instrumentation Engineers (SPIE) Conference Series}, p. 733. 1986.
\bibdoi{10.1117/12.968154}
\bibitem{daophot}
P.~B. {Stetson}.
\newblock {DAOPHOT: A Computer Program for Crowded-Field Stellar Photometry}.
\newblock \textit{\pasp} \textbf{99}:191, 1987.
\bibdoi{10.1086/131977}
\bibitem{sextractor}
E.~{Bertin}, S.~{Arnouts}.
\newblock {SExtractor: Software for source extraction.}
\newblock \textit{\aaps} \textbf{117}:393--404, 1996.
\bibdoi{10.1051/aas:1996164}
\bibitem{astropy}
{Astropy Collaboration}, T.~P. {Robitaille}, E.~J. {Tollerud}, et~al.
\newblock {Astropy: A community Python package for astronomy}.
\newblock \textit{\aap} \textbf{558}:A33, 2013.
\eprint{1307.6212} \bibdoi{10.1051/0004-6361/201322068}
\bibitem{photutils}
L.~{Bradley}, B.~{Sip{\H{o}}cz}, T.~{Robitaille}, et~al.
\newblock {astropy/photutils: 1.0.0}, 2020.
\bibdoi{10.5281/zenodo.4044744}
\bibitem{GRANDMAO3A}
S.~{Antier}, S.~{Agayeva}, V.~{Aivazyan}, et~al.
\newblock {The first six months of the Advanced LIGO's and Advanced Virgo's third observing run with GRANDMA}.
\newblock \textit{\mnras} \textbf{492}(3):3904--3927, 2020.
\eprint{1910.11261} \bibdoi{10.1093/mnras/stz3142}
\bibitem{GRANDMA03B}
S.~{Antier}, S.~{Agayeva}, M.~{Almualla}, et~al.
\newblock {GRANDMA observations of advanced LIGO's and advanced Virgo's third observational campaign}.
\newblock \textit{\mnras} \textbf{497}(4):5518--5539, 2020.
\eprint{2004.04277} \bibdoi{10.1093/mnras/staa1846}
\bibitem{stdpipe}
S.~{Karpov}.
\newblock {STDPipe: Simple Transient Detection Pipeline}.
\newblock Astrophysics Source Code Library, record ascl:2112.006, 2021.

\bibitem{astrometrynet}
D.~{Lang}, D.~W. {Hogg}, K.~{Mierle}, et~al.
\newblock {Astrometry.net: Blind Astrometric Calibration of Arbitrary Astronomical Images}.
\newblock \textit{\aj} \textbf{139}:1782--1800, 2010.
\eprint{0910.2233} \bibdoi{10.1088/0004-6256/139/5/1782}
\bibitem{scamp}
E.~{Bertin}.
\newblock {Automatic Astrometric and Photometric Calibration with SCAMP}.
\newblock In C.~{Gabriel}, C.~{Arviset}, D.~{Ponz}, S.~{Enrique} (eds.), \textit{Astronomical Data Analysis Software and Systems XV}, vol. 351 of \textit{Astronomical Society of the Pacific Conference Series}, p. 112. 2006.

\bibitem{swarp}
E.~{Bertin}.
\newblock {SWarp: Resampling and Co-adding FITS Images Together}.
\newblock Astrophysics Source Code Library, record ascl:1010.068, 2010.

\bibitem{hotpants}
A.~{Becker}.
\newblock {HOTPANTS: High Order Transform of PSF ANd Template Subtraction}.
\newblock Astrophysics Source Code Library, record ascl:1504.004, 2015.

\bibitem{matplotlib}
J.~D. Hunter.
\newblock Matplotlib: A 2d graphics environment.
\newblock \textit{Computing in Science \& Engineering} \textbf{9}(3):90--95, 2007.
\bibdoi{10.1109/MCSE.2007.55}
\bibitem{astroscrappy}
C.~{McCully}, M.~{Tewes}.
\newblock {Astro-SCRAPPY: Speedy Cosmic Ray Annihilation Package in Python}, 2019.
\eprint{1907.032}
\bibitem{astroquery}
A.~{Ginsburg}, B.~M. {Sip{\H{o}}cz}, C.~E. {Brasseur}, et~al.
\newblock {astroquery: An Astronomical Web-querying Package in Python}.
\newblock \textit{\aj} \textbf{157}(3):98, 2019.
\eprint{1901.04520} \bibdoi{10.3847/1538-3881/aafc33}
\bibitem{reproject}
T.~{Robitaille}, C.~{Deil}, A.~{Ginsburg}.
\newblock {reproject: Python-based astronomical image reprojection}.
\newblock Astrophysics Source Code Library, record ascl:2011.023, 2020.

\bibitem{scikit-learn}
F.~Pedregosa, G.~Varoquaux, A.~Gramfort, et~al.
\newblock Scikit-learn: Machine learning in {P}ython.
\newblock \textit{Journal of Machine Learning Research} \textbf{12}:2825--2830, 2011.

\bibitem{lacosmic}
P.~G. {Dokkum (van)}.
\newblock {Cosmic-Ray Rejection by Laplacian Edge Detection}.
\newblock \textit{\pasp} \textbf{113}(789):1420--1427, 2001.
\eprint{astro-ph/0108003} \bibdoi{10.1086/323894}
\bibitem{simbad}
M.~{Wenger}, F.~{Ochsenbein}, D.~{Egret}, et~al.
\newblock {The SIMBAD astronomical database. The CDS reference database for astronomical objects}.
\newblock \textit{\aaps} \textbf{143}:9--22, 2000.
\eprint{astro-ph/0002110} \bibdoi{10.1051/aas:2000332}
\bibitem{tns}
A.~{Gal-Yam}.
\newblock {The TNS alert system}.
\newblock In \textit{American Astronomical Society Meeting Abstracts}, vol. 237 of \textit{American Astronomical Society Meeting Abstracts}, p. 423.05. 2021.

\bibitem{isolationforest}
F.~T. Liu, K.~M. Ting, Z.-H. Zhou.
\newblock Isolation forest.
\newblock In \textit{2008 Eighth IEEE International Conference on Data Mining}, pp. 413--422. 2008.
\bibdoi{10.1109/ICDM.2008.17}
\bibitem{2007PASP..119.1462B}
A.~C. {Becker}, N.~M. {Silvestri}, R.~E. {Owen}, et~al.
\newblock {In Pursuit of LSST Science Requirements: A Comparison of Photometry Algorithms}.
\newblock \textit{\pasp} \textbf{119}(862):1462--1482, 2007.
\eprint{0712.0637} \bibdoi{10.1086/524710}
\bibitem{2013PASP..125...68A}
M.~{Annunziatella}, A.~{Mercurio}, M.~{Brescia}, et~al.
\newblock {Inside Catalogs: A Comparison of Source Extraction Software}.
\newblock \textit{\pasp} \textbf{125}(923):68, 2013.
\eprint{1212.0564} \bibdoi{10.1086/669333}
\bibitem{psfex}
E.~{Bertin}.
\newblock {Automated Morphometry with SExtractor and PSFEx}.
\newblock In I.~N. {Evans}, A.~{Accomazzi}, D.~J. {Mink}, A.~H. {Rots} (eds.), \textit{Astronomical Data Analysis Software and Systems XX}, vol. 442 of \textit{Astronomical Society of the Pacific Conference Series}, p. 435. 2011.

\bibitem{hips2fits}
T.~{Boch}, P.~{Fernique}, F.~{Bonnarel}, et~al.
\newblock {HiPS2FITS: Fast Generation of FITS Cutouts From HiPS Image Datasets}.
\newblock In R.~{Pizzo}, E.~R. {Deul}, J.~D. {Mol}, et~al. (eds.), \textit{Astronomical Data Analysis Software and Systems XXIX}, vol. 527 of \textit{Astronomical Society of the Pacific Conference Series}, p. 121. 2020.

\bibitem{vizier}
F.~{Ochsenbein}, P.~{Bauer}, J.~{Marcout}.
\newblock {The VizieR database of astronomical catalogues}.
\newblock \textit{\aaps} \textbf{143}:23--32, 2000.
\eprint{astro-ph/0002122} \bibdoi{10.1051/aas:2000169}
\bibitem{ps1surveys}
K.~C. {Chambers}, E.~A. {Magnier}, N.~{Metcalfe}, et~al.
\newblock {The Pan-STARRS1 Surveys}.
\newblock \textit{arXiv:161205560} 2016.

\bibitem{skymapper}
C.~A. {Onken}, C.~{Wolf}, M.~S. {Bessell}, et~al.
\newblock {SkyMapper Southern Survey: Data Release 4}.
\newblock \textit{arXiv e-prints} arXiv:2402.02015, 2024.
\eprint{2402.02015} \bibdoi{10.48550/arXiv.2402.02015}
\bibitem{gaiaedr3}
{Gaia Collaboration}, A.~G.~A. {Brown}, A.~{Vallenari}, et~al.
\newblock {Gaia Early Data Release 3. Summary of the contents and survey properties}.
\newblock \textit{AAP} \textbf{649}:A1, 2021.
\eprint{2012.01533} \bibdoi{10.1051/0004-6361/202039657}
\bibitem{refcat2}
J.~L. {Tonry}, L.~{Denneau}, H.~{Flewelling}, et~al.
\newblock {The ATLAS All-Sky Stellar Reference Catalog}.
\newblock \textit{\apj} \textbf{867}(2):105, 2018.
\eprint{1809.09157} \bibdoi{10.3847/1538-4357/aae386}
\bibitem{syntphot}
{Gaia Collaboration}, P.~{Montegriffo}, M.~{Bellazzini}, et~al.
\newblock {Gaia Data Release 3. The Galaxy in your preferred colours: Synthetic photometry from Gaia low-resolution spectra}.
\newblock \textit{AAP} \textbf{674}:A33, 2023.
\eprint{2206.06215} \bibdoi{10.1051/0004-6361/202243709}
\bibitem{Pancino_2022}
E.~{Pancino}, P.~M. {Marrese}, S.~{Marinoni}, et~al.
\newblock {The Gaia EDR3 view of Johnson-Kron-Cousins standard stars: the curated Landolt and Stetson collections}.
\newblock \textit{\aap} \textbf{664}:A109, 2022.
\eprint{2205.06186} \bibdoi{10.1051/0004-6361/202243939}
\bibitem{xmatch}
F.-X. {Pineau}, T.~{Boch}, S.~{Derri{\`e}re}, A.~{Schaaff}.
\newblock {The CDS Cross-match Service: Key Figures, Internals and Future Plans}.
\newblock In P.~{Ballester}, J.~{Ibsen}, M.~{Solar}, K.~{Shortridge} (eds.), \textit{Astronomical Data Analysis Software and Systems XXVII}, vol. 522 of \textit{Astronomical Society of the Pacific Conference Series}, p. 125. 2020.

\bibitem{skybot}
J.~{Berthier}, F.~{Vachier}, W.~{Thuillot}, et~al.
\newblock {SkyBoT, a new VO service to identify Solar System objects}.
\newblock In C.~{Gabriel}, C.~{Arviset}, D.~{Ponz}, S.~{Enrique} (eds.), \textit{Astronomical Data Analysis Software and Systems XV}, vol. 351 of \textit{Astronomical Society of the Pacific Conference Series}, p. 367. 2006.

\bibitem{alard_lupton_1998}
C.~{Alard}, R.~H. {Lupton}.
\newblock {A Method for Optimal Image Subtraction}.
\newblock \textit{\apj} \textbf{503}(1):325--331, 1998.
\eprint{astro-ph/9712287} \bibdoi{10.1086/305984}
\bibitem{zogy}
B.~{Zackay}, E.~O. {Ofek}, A.~{Gal-Yam}.
\newblock {Proper Image Subtraction{\textemdash}Optimal Transient Detection, Photometry, and Hypothesis Testing}.
\newblock \textit{\apj} \textbf{830}(1):27, 2016.
\eprint{1601.02655} \bibdoi{10.3847/0004-637X/830/1/27}
\bibitem{legacysurveys}
A.~{Dey}, D.~J. {Schlegel}, D.~{Lang}, et~al.
\newblock {Overview of the DESI Legacy Imaging Surveys}.
\newblock \textit{\aj} \textbf{157}(5):168, 2019.
\eprint{1804.08657} \bibdoi{10.3847/1538-3881/ab089d}
\bibitem{aladin}
F.~{Bonnarel}, P.~{Fernique}, O.~{Bienaym{\'e}}, et~al.
\newblock {The ALADIN interactive sky atlas. A reference tool for identification of astronomical sources}.
\newblock \textit{\aaps} \textbf{143}:33--40, 2000.
\bibdoi{10.1051/aas:2000331}
\bibitem{ztf_rb}
D.~A. {Duev}, A.~{Mahabal}, F.~J. {Masci}, et~al.
\newblock {Real-bogus classification for the Zwicky Transient Facility using deep learning}.
\newblock \textit{\mnras} \textbf{489}(3):3582--3590, 2019.
\eprint{1907.11259} \bibdoi{10.1093/mnras/stz2357}
\bibitem{otrain}
K.~{Makhlouf}, D.~{Turpin}, D.~{Corre}, et~al.
\newblock {O'TRAIN: A robust and flexible `real or bogus' classifier for the study of the optical transient sky}.
\newblock \textit{\aap} \textbf{664}:A81, 2022.
\eprint{2112.10280} \bibdoi{10.1051/0004-6361/202142952}
\bibitem{autophot}
S.~J. {Brennan}, M.~{Fraser}.
\newblock {The Automated Photometry of Transients pipeline (AUTOPHOT)}.
\newblock \textit{\aap} \textbf{667}:A62, 2022.
\eprint{2201.02635} \bibdoi{10.1051/0004-6361/202243067}
\bibitem{photometrypipeline}
M.~{Mommert}.
\newblock {PHOTOMETRYPIPELINE: An automated pipeline for calibrated photometry}.
\newblock \textit{Astronomy and Computing} \textbf{18}:47--53, 2017.
\eprint{1702.00834} \bibdoi{10.1016/j.ascom.2016.11.002}
\bibitem{vast}
K.~V. {Sokolovsky}, A.~A. {Lebedev}.
\newblock {VaST: A variability search toolkit}.
\newblock \textit{Astronomy and Computing} \textbf{22}:28--47, 2018.
\eprint{1702.07715} \bibdoi{10.1016/j.ascom.2017.12.001}
\bibitem{astroimagej}
K.~A. {Collins}, J.~F. {Kielkopf}, K.~G. {Stassun}, F.~V. {Hessman}.
\newblock {AstroImageJ: Image Processing and Photometric Extraction for Ultra-precise Astronomical Light Curves}.
\newblock \textit{\aj} \textbf{153}(2):77, 2017.
\eprint{1701.04817} \bibdoi{10.3847/1538-3881/153/2/77}
\bibitem{apt}
R.~R. {Laher}, V.~{Gorjian}, L.~M. {Rebull}, et~al.
\newblock {Aperture Photometry Tool}.
\newblock \textit{\pasp} \textbf{124}(917):737, 2012.
\bibdoi{10.1086/666883}
\bibitem{fink}
A.~{M{\"o}ller}, J.~{Peloton}, E.~E.~O. {Ishida}, et~al.
\newblock {FINK, a new generation of broker for the LSST community}.
\newblock \textit{\mnras} \textbf{501}(3):3272--3288, 2021.
\eprint{2009.10185} \bibdoi{10.1093/mnras/staa3602}
\end{thebibliography}

\end{document}